\documentclass{andp2012}
\usepackage{graphicx}

\newcommand{\Tr}{\mathrm{Tr}}
\newcommand{\cd}{\hat{c}^\dagger}
\renewcommand{\c}{\hat{c}^{\phantom\dagger}}

\newcommand{\taud}{\hat{\tau}^{\dagger}}
\newcommand{\taua}{\hat{\tau}^{\phantom\dagger}}
\newcommand{\tauz}{\hat{\tau}^{z}}

\title{Many-body localization and delocalization from the perspective of Integrals of Motion}
\author[L. Rademaker]{Louk Rademaker\inst{1,}\footnote{Corresponding author\quad E-mail:~\textsf{louk.rademaker@gmail.com}}}
\author[M. Ortu\~{n}o Author]{Miguel Ortu\~{n}o\inst{2}}
\author[A. M. Somoza]{Andres M. Somoza\inst{2}}
\address[1]{Kavli Institute for Theoretical Physics, University of California Santa Barbara, CA 93106, USA}
\address[2]{Departamento de F\'{i}sica - CIOyN, Universidad de Murcia, Murcia 30.071, Spain}
\shortauthors{L. Rademaker et al.}
\begin{abstract}
  We study many-body localization (MBL) and delocalization from the perspective of integrals of motion (IOMs). MBL can be understood phenomenologically through the existence of macroscopically many localized IOMs. However, IOMs exist for all many-body systems, and non-localized IOMs determine properties on the ergodic side of the MBL transition too. Here we explore their properties using our method of displacement transformations. We show how different quantities can be calculated using the IOMs as an expansion in the number of operators. For all values of disorder the typical IOMs are localized, suggesting the importance of rare fluctuations in understanding the delocalization transition.
\end{abstract}
\shortabstract
\begin{document}
\maketitle

\noindent Any quantum many-body system has as many conserved quantities as degrees of freedom. Historically this observation was considered trite and inconsequential, for the simple reason that these integrals of motion (IOMs) are too complicated to be of practical relevance. When the conserved quantities turn out to be accessible, whether trivially as in noninteracting systems or only after profound leaps such as the Bethe ansatz, we speak of an `integrable' system. However, `non-integrable' systems do have IOMs too that will constrain both dynamical as well as statistical properties.

Recently, IOMs gained a renewed interest in the context of interacting systems with disorder. In the non-interacting Anderson insulator\cite{Anderson:1958fz} in $d=1,2$ dimensions the single particle wavefunctions are exponentially localized. Even in the presence of weak interactions particles remain localized, which is known as many-body localization (MBL).\cite{Basko:2007ta,Basko:2006vz,Nandkishore:2015kt} Recently, it was realized that MBL can be understood through the existence of an extensive number of exponentially localized IOMs.\cite{2014arXiv1407.8480C,Huse:2014co,Serbyn:2013cl} Inevitably, this observation led to a rush of new methods to compute the IOMs in the MBL-phase,\cite{2014arXiv1412.3073K,2015NuPhB.891..420R,Imbrie2016,YouQiXu2016,2016arXiv160707884P,
2016arXiv160801328G,
2016arXiv160803296O,
2016arXiv160300440I,Friesdorf:2015dd,2016arXiv160609509H} and recently we published our own computational method using displacement transformations.\cite{Rademaker:2016jf}

The presence of these localized IOMs in the fully many-body localized phases prevents thermalization. The question of whether a many-body quantum system thermalizes has been cast into the Eigenstate Thermalization Hypothesis (ETH):\cite{Deutsch:1991ju,1996JPhA...29L..75S,Srednicki:1994dl,Rigol:2011bf} the expectation value of any local observable in an eigenstate with a given energy density is equal to its expectation value in the Gibbs ensemble with corresponding temperature. If, however, there exist local density IOMs (that can be expressed as the sum of local operators) the corresponding thermal state will be a so-called generalized Gibbs ensemble.\cite{Rigol:2006df} It has been shown that any finite-ranged translationally invariant Hamiltonian will thermalize towards their corresponding generalized Gibbs ensemble.\cite{2015arXiv151203713D} Therefore, whether and how a system thermalizes is directly related to the structure of its IOMs.

However, there are only two cases where the Hamiltonian of the system is commonly written out in terms of the IOMs. One case is Fermi liquid theory,\cite{Pines:1999wg} where
the energy is written as $E_{FL} = \sum_p \xi_p n_p + \frac{1}{2} \sum_{pp'} f_{pp'} n_p n_{p'} + \ldots$, where $n_p$ are classical occupation numbers of quasiparticles with momentum $p$ - which are nothing other than IOMs! The other case is the MBL phase, where the following effective classical Hamiltonian was proposed\cite{Nandkishore:2015kt}
\begin{equation}
	\hat{H} = \sum_i \xi_i \hat{\tau}^z_i 
		+ \sum_{i<j} J_{ij} \hat{\tau}^z_i \hat{\tau}^z_j 
		+ \sum_{i<j<k} J_{ijk} \hat{\tau}^z_i \hat{\tau}^z_j \tauz_k + \ldots,
	\label{LIOM}
\end{equation}
where $\tauz_i$ are the IOMs. The question is whether one can write a classical Hamiltonian \`{a} la Eqn. (\ref{LIOM}) for \emph{any} many-body system, specifically, also for the ergodic phase of disordered interacting systems?

In Sec. \ref{Sec1GeneralIOMs} we will discuss some general aspects of expressing many-body systems in terms of their IOMs. We will show that indeed, formally, one can write any interacting system - even 'non-integrable' ones - into the classical form of  Eqn. (\ref{LIOM}). On a practical level, we develop a systematic way to construct the IOMs. Thereby we construct many-body states that are generalizations of Slater determinant product states. The computational complexity is thereby reduced to that of solving exactly few-body problems. As a brief aside we discuss the relation between the IOMs introduced above and the traditional notion of integrability.


Subsequently, in Sec. \ref{Sec2Results} we apply these methods to the problem of the delocalization transition in a one dimensional interacting disordered system. Our model is the Anderson insulator of spinless fermions with nearest-neighbor repulsion, which is equivalent to the Heisenberg chain with random fields. This model is known to exhibit a $T=\infty$ transition from an MBL phase at large disorder, to an ergodic phase at small disorder.\cite{Znidaric:2008cr,Pal:2010gr,Bardarson:2012gc,DeLuca:2013ba,Bauer:2013jw} However, it appears that the typical properties of the IOMs do not reflect this transition. Finally, we will discuss possible ways how the delocalization transition can be understood from the perspective of IOMs.

\section{General remarks on the IOM-basis}
\label{Sec1GeneralIOMs}

In this section we will discuss how to write a Hamiltonian in terms of its IOMs. We will consider a general model of interacting fermions on a lattice with $N$ sites, that preserves the total number of fermions $\hat{n}_{\mathrm{tot}} = \sum_i \hat{n}_i$. The Hamiltonian of such a system can be written as
\begin{equation}
	\hat{H} = \sum_i \xi_i \hat{n}_i 
	+ \frac{1}{2} \sum_{i jkl} V_{ijkl} 
		\cd_i \cd_j \c_k \c_l
	+ \ldots
	\label{InterCBasis}
\end{equation}
where $\hat{n}_i = \hat{c}^\dagger_i \hat{c}_i^{\phantom\dagger}$ is the number operator and $\{ \hat{c}^\dagger_i, \hat{c}_j^{\phantom\dagger} \} = \delta_{ij}$ is the standard anticommutation condition on the fermion creation and annihilation operators. All operators are denoted with a hat. Note that for convenience we have, in the above notation, already diagonalized the quadratic part of the Hamiltonian. Furthermore, the labels $i$ indicate any type of quantum numbers, which can be either real space or momentum space, and can possibly include spin and orbital degrees of freedom. We will refer to the basis in which Eqn. (\ref{InterCBasis}) is written as the {\em original basis}.

The central claim of this section is that there exists a unitary transformation $\hat{U}$ such that $\hat{H}' = \hat{U}^\dagger \hat{H} \hat{U}$ is classical, following Eqn. (\ref{LIOM}),
\begin{equation}
	\hat{H}' = \sum_i \xi_i \hat{\tau}^z_i + \sum_{i<j} J_{ij} \hat{\tau}^z_i \hat{\tau}^z_j + 
	\sum_{i<j<k} J_{ijk}  \hat{\tau}^z_i \hat{\tau}^z_j \hat{\tau}^z_k + \ldots.
	\label{LIOM2}
\end{equation}
We choose the $N$ integrals of motion $\tauz_i$ to be number operators, so that they have only eigenvalues zero and one. Since they are IOMs they commute with each other and with the Hamiltonian,
\begin{equation}
	[ \hat{\tau}^z_i, \hat{\tau}^z_j ] = [ \hat{\tau}^z_i, \hat{H}] = 0.
	\label{Commutator}
\end{equation}
We will call the basis of the IOMs the {\em classical} or {\em $\tau$-basis}. Below we will show that any many-body eigenstate is a product state in the $\tau$-basis, $| \psi_n \rangle = \taud_{i_1} \cdots \taud_{i_k} | 0 \rangle$.

In Sec. \ref{SSecFormalConstr} we will provide a formal construction as proof that indeed the $\tau$-basis exists, in spirit similar to Ref. \cite{Lychkovskiy:2013bs}. In Sec. \ref{SecPractical} we show that through consecutive applications of displacement transformations the classical basis can be computed. Furthermore, we propose that an approximate form of the transformation $\hat{U}$ can be constructed by clever use of few-particle exact states. Because the classical basis is by no means unique, in Sec. \ref{SecNonUnique} we discuss how one can find the best choice of IOMs. Finally, we briefly address in Sec. \ref{SecIntegrable} the relation between the $\tau$-basis and the field of integrable quantum systems.

\subsection{Formal construction}
\label{SSecFormalConstr}

Consider an interacting number-conserving fermion Hamiltonian on a lattice with $N$ sites, for example Eqn. (\ref{InterCBasis}). The associated Hilbert space $\mathcal{H}$ is $2^N$ dimensional, and can be split into $N+1$ subspaces of fixed particle number,
\begin{equation}
	\mathcal{H} = \otimes_{k=0 \ldots N} \mathcal{H}^{(k)}.
\end{equation}
Here $\mathcal{H}^{(k)}$ is the $\binom{N}{k} $-dimensional subspace containing states with $k$ particles. 

Let us label all the eigenstates of the Hamiltonian $\hat{H}$ as $| k, n \rangle$ where $k = 0, \ldots, N$ labels the number of particles and $n = 1, \ldots, \binom{N}{k}$ is an index enumerating the eigenstates within the $k$-particle subspace. The projection operator onto the eigenstate $|k,n \rangle$ is $\hat{P}^{(k)}_n = |k,n \rangle \langle k,n |$. The Hamiltonian can be written as
\begin{equation}
	\hat{H} = \sum_{k=0}^N \sum_{n=1}^{\binom{N}{k}} \hat{P}^{(k)}_n E_{k,n} 
	\label{Hcrazy}
\end{equation}
where $E_{k,n}$ are the eigenvalues of the Hamiltonian. Even though this is a basis where the Hamiltonian is expressed in terms of  $2^N$ integrals of motion, it is not the desired form of Eqn. (\ref{LIOM}).

Instead, we wish to construct a set of $N$ projection operators $\hat{\tau}^z_j$ with $j = 1, \ldots, N$. This can be done in each $k$-particle subspace separately, starting with the single particle subspace $\mathcal{H}^{(1)}$. We equate the $\hat{\tau}^z_j$ operators with the projectors onto the eigenstates,
\begin{equation}
	\left. \hat{\tau}^z_j \right|_{\mathcal{H}^{(1)}} = \hat{P}^{(1)}_j.
\end{equation}
This is trivially possible, since there are $N$ single-particle eigenstates and an equal number of $\hat{\tau}^z$-operators. Similarly, within the $k$-particle subspace $\mathcal{H}^{(k)}$ we equate $\hat{\tau}^z_j$ with the sum over $\frac{k}{N} \binom{N}{k}$ projectors onto eigenstates, such that the product of $k$ $\hat{\tau}^z_j$-operators corresponds to a unique eigenstate projector $\hat{P}^{(k)}_i$,
\begin{equation}
	\hat{P}^{(k)}_n = \left. \hat{\tau}^z_{i_{n,1}} \ldots \hat{\tau}^z_{i_{n,k}} \right|_{\mathcal{H}^{(k)}}.
\end{equation}
This is possible because there are $\binom{N}{k}$ projectors and the same number of unique combinations of $k$ $\hat{\tau}^z$-operators.

As an example, let us write out the mapping for $N=4$ sites and $k=2$ particles. There are $\binom{N}{k}=6$ eigenstates in this subspace, and the projectors onto eigenstates are labelled $\hat{P}^{(2)}_n$ with $n=1, \ldots, 6$. There are $N=4$ of the $\hat{\tau}^z$-operators, and each is equal to the sum over $\frac{k}{N} \binom{N}{k} = 3$ eigenstate projectors. A possible matching is
\begin{eqnarray*}
	\left. \hat{\tau}^z_1  \right|_{\mathcal{H}^{(2)}}& = & \hat{P}^{(2)}_1 + \hat{P}^{(2)}_2 + \hat{P}^{(2)}_3, \\
	\left. \hat{\tau}^z_2  \right|_{\mathcal{H}^{(2)}}& = & \hat{P}^{(2)}_1 + \hat{P}^{(2)}_4 + \hat{P}^{(2)}_5, \\
	\left. \hat{\tau}^z_3  \right|_{\mathcal{H}^{(2)}}& = & \hat{P}^{(2)}_2 + \hat{P}^{(2)}_4 + \hat{P}^{(2)}_6, \\
	\left. \hat{\tau}^z_4  \right|_{\mathcal{H}^{(2)}}& = & \hat{P}^{(2)}_3 + \hat{P}^{(2)}_5 + \hat{P}^{(2)}_6.
\end{eqnarray*}
Now every projector onto an eigenstate in the $k=2$ space is uniquely the product of two $\hat{\tau}^z$-operators, for example $\hat{P}^{(2)}_1 = \left. \hat{\tau}^z_1 \hat{\tau}^z_2 \right|_{\mathcal{H}^{(2)}}$. 

Going back to the general case, we make such a mapping for each $k$-particle subspace up to the $N$-particle space, where the projector onto the $N$-particle state is the product of all $\tauz$-operators, $\hat{P}^{(N)} = \left. \prod_{i}^N \tauz_i \right|_{\mathcal{H}^{(N)}}$.

We have thus constructed $N$ projection operators $\hat{\tau}^z_i$ such that every eigenstate projector $P^{(k)}_n$ can be written as a product of $\hat{\tau}^z$-operators. Therefore the Hamiltonian Eqn. (\ref{Hcrazy}) is transformed into the desired classical form,
\begin{equation}
	\hat{H}' = \sum_i \xi_i \hat{\tau}^z_i 
	+ \sum_{i<j} J_{ij} \hat{\tau}^z_i \hat{\tau}^z_j 
	+ \sum_{i<j<k} J_{ijk} \hat{\tau}^z_i \hat{\tau}^z_j  \hat{\tau}^z_k +  \ldots.
	\label{Diagonal}
\end{equation}
Furthermore, because the eigenstate projection operators commute with each other and with the Hamiltonian, it follows that $\tauz$ operators are also IOMs, as in Eqn. (\ref{Commutator}). Note that the set of integrals of motion $(\tauz_1 , \ldots , \tauz_N)$ thus constructed is algebraically independent.

In the basis thus introduced it follows that every eigenstate of the Hamiltonian $\hat{H}$ is uniquely specified by the eigenvalue of each $\tauz$-operator,
\begin{equation}
	| k, n \rangle = | \tau^z_1 \cdots \tau^z_N \rangle
	\label{TauBasis-State}
\end{equation}
where $\tau^z_i = 0,1$ are the eigenvalues of this state when acted upon by $\hat{\tau}^z_i$. In our $N=4$ example mentioned before, the state $|k=2,n=1\rangle$ corresponds to the state where $\tau^z_1$ and $\tau^z_2$ are occupied, hence $|k=2,n=1 \rangle = | 1100\rangle$.

Acting alongside the $\tauz$-operators, there exist $N$ fermion creation operators $\hat{\tau}^\dagger_j$, with $j = 1, \ldots, N$, anticommutation relation $\{ \taud_i, \taua_j \} = \delta_{ij}$ and $[ \hat{\tau}^z_i , \hat{\tau}^\dagger_j ] = \delta_{ij} \hat{\tau}^\dagger_i$, with $\tauz_i = \taud_i \taua_i$ such that every $k$-particle eigenstate $|k, n \rangle$ can be written as
\begin{equation}
	|k,n \rangle = \hat{\tau}^\dagger_{j_1} \ldots \hat{\tau}^\dagger_{j_k} | 0 \rangle,
	\label{GenProdState}
\end{equation}
where $|0 \rangle$ is the vacuum state without particles. In other words, \emph{every eigenstate of $\hat{H}$ is a product state} in the classical $\tau$-basis. States of the form dictated by Eqn. (\ref{GenProdState}) in the $\tau$-basis are called \emph{generalized product states}.

The creation operators for a $\tau$-state, that is $\hat{\tau}^\dagger_i$, can be explicitly constructed using the mapping provided by Eqn. (\ref{TauBasis-State}),
\begin{eqnarray}
	\hat{\tau}^\dagger_i &=& \sum_{j \neq i, \tau^z_j = 0,1} (-1)^{\sum_{k<i} \tau^z_k} 
	 \\ &&	
	| \cdots \tau^z_{i-1} (\tau^z_i = 1) \tau^z_{i+1} \cdots \rangle
		\langle \cdots \tau^z_{i-1} (\tau^z_i = 0) \tau^z_{i+1} \cdots |.
		\nonumber
\end{eqnarray}
The minus sign is to ensure anticommutation relations between the $\hat{\tau}^\dagger_i$ operators. If one does not include the minus signs, the resulting operators satisfy an algebra of hard-core bosons $\hat{b}^\dagger_i$. The bosonic operators $\hat{b}^\dagger_i$ are related to the fermionic $\hat{\tau}^\dagger_i$ via the Jordan-Wigner transformation.

In the above considerations we have explicitly discussed a system with $N$ sites. A natural question is whether this construction is still valid in the thermodynamic limit $N \rightarrow \infty$. At strictly $N= \infty$, the notion of an eigenstate becomes ill-defined and naturally the $\tau$-basis cannot be formulated. However, for any finite $N$, no matter how large, we can construct the $\tau$-basis. Therefore we will adhere to a practical assessment of the thermodynamic limit, that is: if computed quantities converge with increasing $N$ we consider the $\tau$-basis to be valid in the thermodynamic limit.

\subsection{Practical method}
\label{SecPractical}

We have shown in Sec. \ref{SSecFormalConstr} that there exists a classical $\tau$-basis for each interacting Hamiltonian. Therefore there exists a unitary transformation $\hat{U}$ that brings us from the original basis to the classical basis:
\begin{equation}
	\hat{H}' = \hat{U}^\dagger \hat{H} \hat{U}, \; \;
	\tauz_i =\hat{U}^\dagger \hat{n}_i \hat{U}, \; \;
	\taud_i = \hat{U}^\dagger \cd_i \hat{U}.
\end{equation}
Because the product of unitary operators is unitary again, we can divide the full transformation $\hat{U}$ into a product of `simple' transformations $\hat{\mathcal{D}}_i$
\begin{equation}
	\hat{U} = \hat{\mathcal{D}}_1 \hat{\mathcal{D}}_2 \ldots \hat{\mathcal{D}}_M
\end{equation}
Now any unitary transformation can be written as the exponential of an anti-Hermitian matrix. Furthermore, any anti-Hermitian matrix can be written as $\hat{X}^\dagger - \hat{X}$ for some operator $\hat{X}$. Consequently, we express each 'simple' operator as
\begin{equation}
	\hat{\mathcal{D}}_i = \exp \left( \lambda_i ( \hat{X}^\dagger_i - \hat{X}_i ) \right).
\end{equation}
and for `simplicity' we require $\hat{X}$ to satisfy the following properties,
\begin{equation}
	\hat{X}^2 = 0, \; \mathrm{and} \; \hat{X}\hat{X}^\dagger\hat{X} = \hat{X}.
\end{equation}
These properties allow us to expand the `simple' transformation exactly,
\begin{equation}
	\hat{\mathcal{D}}_i = 1 + \sin \lambda_i (\hat{X}^\dagger_i - \hat{X}_i ) 
		+ (\cos \lambda_i - 1) ( \hat{X}^\dagger \hat{X} + \hat{X} \hat{X}^\dagger).
\end{equation}
These are the {\em displacement transformations} that we introduced in our previous work.\cite{Rademaker:2016jf} There we used subsequent applications of displacement transformations to diagonalize the Hamiltonian. Such iterative procedure can be viewed as a discrete version of the Wegner flow equations, who recently have been applied to study many-body localization.\cite{2001PhR...348...77W,Monthus2016,2016arXiv160707884P,2016arXiv160603094Q}

For a Hamiltonian of the form Eqn. (\ref{InterCBasis}), we first performed displacement transformations where the operator $\hat{X}$ is the product of two creation and two annihilation operators, $\hat{X} = \cd_i \cd_j \c_k \c_l$. Transformations of this type allow us to make the Hamiltonian classical up to and including all four-operator terms,
\begin{equation}
	\hat{H}' = \xi_i \hat{n}_i + J_{ij} \hat{n}_i \hat{n}_j 
		+ V_{i_1 i_2 i_3 i_4 i_5 i_6} \cd_{i_1}\cd_{i_2}\cd_{i_3}\c_{i_4}\c_{i_5}\c_{i_6} + \ldots
	\label{Hfourth}
\end{equation}
provided that all higher-order terms in the Hamiltonian are normal-ordered with respect to the particle vacuum. Note that in the above equation the $\cd_i$ and $\hat{n}_i$ operators are not the same as in the original basis - they are transformed by all the fourth order displacement transformations.
We can multiply the displacement transformations with $\hat{X}$ containing four operators into one transformation $\hat{U}_4$ that brings the Hamiltonian into the form of Eqn. (\ref{Hfourth}). Similarly, we can then construct a transformation $\hat{U}_6$ that brings the sixth-order part into a classical form, where now we have grouped together all displacement transformations where $\hat{X}$ is the product of six fermionic operators.

This appears at first sight to be a useless rewriting of displacement transformations. However, realizing there exists a transformation $\hat{U}_4$ that brings the original Hamiltonian Eqn. (\ref{InterCBasis}) into Eqn. (\ref{Hfourth}) allows us to find a short-cut to compute $\hat{U}_4$. Consider the space of two-particle states, $|i j \rangle = \cd_i \cd_j | 0\rangle$. Because of the normal ordering, the sixth and higher order terms in Eqn. (\ref{Hfourth}) do not act on the two-particle space. Therefore Eqn. (\ref{Hfourth}) is diagonal in the two-particle space, and the transformation matrix $\hat{U}_4$ when restricted to the 2-particle space equals the transformation $\hat{\mathcal{U}}_2$ that diagonalizes the two-particle spectrum. Thus we write
\begin{equation}
	\hat{U}_4 = \exp \left[ \mathcal{A}_{ij}\,^{kl} \cd_i \cd_j \c_l \c_k \right]
	\label{U4Eqn}
\end{equation}
and interpreting $\mathcal{A}$ as a $\binom{N}{2}$-dimensional matrix, we find the matrix elements of $\hat{U}_4$
\begin{eqnarray}
	\langle ij | \hat{U}_4 | kl \rangle &=& \delta_{ij = kl} + \mathcal{A}_{ij}\,^{kl} + \frac{1}{2} (\mathcal{A})^2_{ij}\,^{kl} + \ldots \\
	&=& ( \exp \mathcal{A} )_{ij}\,^{kl} \\
	& =& \langle ij | \hat{\mathcal{U}}_2 | kl \rangle.
	\label{U4Eqn2}
\end{eqnarray}
By solving the exact two-particle spectrum, we can construct a unitary transformation $\hat{U}_4$ that brings the Hamiltonian into the form of Eqn. (\ref{Hfourth})! This yields the same result as doing the displacement transformations at $4th$ order sequentially, however, it is significantly faster.

Subsequently, starting from Eqn. (\ref{Hfourth}), we can solve the three-particle spectrum exactly and construct the transformation $\hat{U}_6 = \exp \left[ \mathcal{A}_{ijk}\,^{lmn}\cd_i\cd_j\cd_k\c_n\c_m\c_l \right]$ where $\exp \mathcal{A}$ equals the unitary matrix that diagonalizes the three-particle space. 

The method we thus propose consists of solving few-particle states exactly, and then extracting from that a unitary transformation that acts on the whole spectrum. The many-body eigenstates are approximated applying $\hat{U}_4$ onto a non-interacting eigenstate,
\begin{equation}
	| \psi_n \rangle = e^{\mathcal{A}_{ij}\,^{kl} \cd_i \cd_j \c_l \c_k} \cd_{i_1} \cdots \cd_{i_k} | 0 \rangle.
\end{equation}
In the above form, our procedure seems to be related to Hartree-Fock methods, since there many-body eigenstates are approximated by $|\psi_{HF} \rangle = e^{\mathcal{A}_{i}\,^{j} \cd_i \c_j } \cd_{i_1} \cdots \cd_{i_k} | 0 \rangle.$

In Sec. \ref{Sec2Results} of this paper we will use the method of few-particle exact diagonalization to compute the parameters $J_{ij}$ and $J_{ijk}$ of the classical Hamiltonian of interacting disordered chains. However, one must be careful when constructing the exact few-particle eigenstates. The matrix $\hat{\mathcal{U}}_2$ is not unique, and in the next subsection we will address how to choose the optimal shape of $\hat{\mathcal{U}}_2$.

\subsection{Basis optimization}
\label{SecNonUnique}

We have thus far shown that there exists a classical $\tau$-basis, and that we can approximate the $\tau$-basis either by displacement transformations or via exact diagonalization of few-particle states. The question is whether this limitation to few-body states is a sensible physical approximation. In this section we will show that the approximations validity can be quantified by the \emph{quasiparticle weight} $Z_i$. Before we can introduce this concept, we need to show that the $\tau$-basis is not unique and can be changed in nontrivial ways.

The best way to show that the $\tau$-basis is not unique is by explicitly introducing a nontrivial transformation relating two $\tau$-bases. Start with a Hamiltonian in a given $\tau$-basis, with parameters as in Eqn. (\ref{Diagonal}). The simplest nontrivial transformation swaps two states, say $\taud_i \taud_j |0 \rangle$ and $\taud_k \taud_l |0\rangle$. Under such a swap the eigenvalues of the model remain the same, yet the parameter in the Hamiltonian change according to
\begin{eqnarray}
	J_{ij}^{\mathrm{new}} &= & \xi_{k} + \xi_l - \xi_i - \xi_j + J_{kl}, \\
	J_{kl}^{\mathrm{new}} &= & \xi_{i} + \xi_j - \xi_k - \xi_l + J_{ij}.
\end{eqnarray}
and 
\begin{equation}
	J_{ijm_1 \ldots m_k} \leftrightarrow J_{klm_1 \ldots m_k}.
\end{equation}
This swap can be written beautifully in terms of a displacement transformation with angle $\lambda = \pi/2$,
\begin{equation}
	\hat{D} = \exp \left[ \frac{\pi}{2} \left( \cd_i \cd_j \c_l \c_k - \mathrm{h.c.} \right) \right].
	\label{PermutationDT}
\end{equation}
Transforming a Hamiltonian that only contains density terms with $\hat{D}$ will only yield density terms, since any new non-density terms generated will have $\sin 2 \lambda = 0$ as prefactor. In general, any displacement transformation with $\lambda = \pi/2$ constitutes a transformation between different $\tau$-bases.

As an example of how one spectrum can be represented by two different classical Hamiltonians, consider the free fermion system with $N=3$ sites
\begin{equation}
	\hat{H} = \sum_{i=1}^3 \xi_i \hat{n}_i.
	\label{H3}
\end{equation}
If we transform this Hamiltonian with the operator $\hat{D} = \exp \frac{\pi}{2} \left( \cd_1 \hat{n}_3 \c_2 - h.c. \right)$ we find
\begin{eqnarray}
	\hat{H} &=& \sum_{i=1}^3 \xi_i \hat{\tau}^z_i 
		+ \left( \xi_1 - \xi_2 \right)  \hat{\tau}^z_2  \hat{\tau}^z_3
		+ \left( \xi_2 - \xi_1 \right)  \hat{\tau}^z_1  \hat{\tau}^z_3
	\label{Hnew3}
\end{eqnarray}
which has clearly different parameters, yet has the same spectrum and eigenspaces.

Given that the diagonal $\tau$-basis is not unique, we can ask which of the possible $\tau$-bases is the 'best'. An optimal basis should be simple and clear, but should also reflect physical properties as best as possible. Can we quantify such simplicity? There are three natural ways to define an optimal basis, which we will now introduce.

The \emph{information-optimal} basis minimizes the number of parameters $V_{i_1 \ldots i_k}$ that are nonzero. In the example above, Eqn. (\ref{Hnew3}) has 5 nonzero parameters whereas the equivalent Eqn. (\ref{H3}) has only 3 nonzero parameters. Obviously, the latter basis is preferable in that we are able to express the same system with less information.

For the \emph{maximally local} or \emph{natural} basis, the support of each integral of motion $\tauz_i$ is centered around site $i$, where `sites' are defined in the original basis. The integrals of motion $\tauz_i$ can be expressed in the original basis as\footnote{If the original Hamiltonian is diagonal at the quadratic level, the second term $\alpha_{i;jk} \cd_i \c_k$ is necessarily absent.}
\begin{equation}
	\tauz_i = \hat{U} \hat{n}_i \hat{U}^\dagger
		= \hat{n}_i + \alpha_{i;jk} \cd_i \c_k + \alpha_{i; jklm} \cd_i \cd_j \c_k \c_l + \ldots
	\label{TauBasisExpansion}
\end{equation}
Now each parameter $\alpha_{i;j_1 \ldots j_k}$ has an associated `distance' defined as the maximum of $|i - j_k|$. In any local basis, the average absolute value of these parameters should decay with distance; and the maximally local basis is where the decay length is the shortest. Note that for an MBL system the local basis is measured in real-space, whereas in a Fermi liquid we require localization in momentum space.

Naturally, in such a maximally local basis the $c$-basis is also `close' to the $\tau$-basis, in a way that we will quantify order by order in the number of particles $k$. The unitary transformation $\hat{U}_4$, defined in Eqn. (\ref{U4Eqn}), measures the overlap between free two-particle states and exact two-particle eigenstates,
\begin{equation}
	\left( \hat{U}_4 \right)_{\alpha_i \alpha_j}\,^{ij}
	= \langle 0 | \taua_{\alpha_j} \taua_{\alpha_i} \; \cd_i \cd_j | 0 \rangle.
\end{equation}
where $\cd_i$ are the operators that create single-particle eigenstates. Through permutations on the rows of $\hat{U}_4$ (that is, permute the $\alpha_i, \alpha_j$-labels) we can maximize the diagonal matrix elements of $\hat{U}_4$. This implies we make each exact two-particle eigenstate to have maximal overlap with free two-particle states. In practice we maximize $\Tr\, \hat{U}_4$.

Since $\cd_i$ are the single-particle eigenstate operators, we find that $\taud_i | 0 \rangle = \cd_i | 0 \rangle$. The matrix elements of $\hat{U}$ can thus similarly be written as
\begin{equation}
	\left( \hat{U}_4 \right)_{\alpha_i \alpha_j}\,^{ij}
	= \langle \psi^{(1)}_{\alpha_j} | \taua_{\alpha_i} \; \cd_i | \psi^{(1)}_{j}\rangle
		\label{U4exp}
\end{equation}
where $| \psi^{(1)}_j \rangle$ is the $j$-th single-particle eigenstate. This notation allows a straightforward generalization to higher order transformations, 
\begin{equation}
	\left( \hat{U}_{2k} \right)_{\alpha_{i_1} \cdots \alpha_{i_{k}}}\,^{i_1 \cdots i_{k}}
	= \langle \psi^{(n-1)}_{\alpha_{i_{k-1}} \cdots \alpha_{i_{1}}} | 
		\taua_{\alpha_k} \; \cd_n | \psi^{(k-1)}_{i_1 \cdots i_{k-1}}\rangle
	\label{U6exp}
\end{equation}
where $| \psi^{(k-1)}_{i_1 \cdots i_{k-1}}\rangle$ represent the $(k-1)$-particle eigenstates, which are product states of the $\cd_{i}$ operators. At any order $k$ we will maximize the trace $\Tr \, \hat{U}_{2n}$.

The Eqns. (\ref{U4exp})-(\ref{U6exp}) are related to the quasiparticle weight as defined in Fermi liquid theory,\cite{Pines:1999wg}
\begin{equation}
	Z_{\ell,n} = | \langle \psi^{(k)}_{n'} | \cd_\ell | \psi_n^{(k-1)} \rangle | ^2 
	\label{QPoverlap}
\end{equation}
which measures the overlap between an exact $k$-particle eigenstate and the state created by adding one electron to an $(k-1)$-particle eigenstate $| \psi_n^{(k-1)} \rangle$. Once we have maximized the trace of $\hat{U}_{2k}$, the square of the diagonal elements of that matrix correspond to the physically relevant quantity $Z_{\ell,n}$ since then $| \psi^{(k)}_{n'} \rangle = \taud_{\ell} | \psi_n^{(k-1)} \rangle$. Notice, that the quasiparticle weight depends on both the electron quantum number $\ell$ as well as the chosen state labeled by $n$. In typical Fermi liquid theory considerations, one computes the quasiparticle weight with respect to the many-body ground state.

The idea to quantify the many-body-localized phase by a quasiparticle weight has been explored earlier in Ref. \cite{Bera:2015jh}. There they diagonalized the single-particle density matrix $\rho_{ij} = \langle \psi_n | \cd_i \c_j | \psi_n \rangle$. For half-filled states in the MBL phase, half of the eigenvalues of $\rho_{ij}$ are close to one, and half are close to zero. The discontinuous jump in the eigenvalue spectrum of $\rho_{ij}$ is associated with the quasiparticle weight $Z$, as is common in Fermi liquid theory.\cite{Pines:1999wg} With this identification, the jump $Z$ in the spectrum of $\rho_{ij}$ is equal to the overlap defined in Eqn. (\ref{QPoverlap}), and is thus equivalent to the diagonal elements of the optimized $\hat{U}_{2k}$.

In the next section we will construct the natural $\tau$-basis using the maximization of the diagonal part of $\hat{U}_4$ and $\hat{U}_6$. 

\subsection{Relation to integrability}
\label{SecIntegrable}

Several quantum systems, amongst them the one-dimen-sional Heisenberg chain, are considered to be `integrable'. Hand-wavingly, this classification implies the existence of macroscopically many conserved charges. Since we have just stated that \emph{all} quantum systems have extensively many IOMs, it is necessary to clarify the relation between the integrable systems and the classical basis of Eqn. (\ref{LIOM}). 

We find that the definition of integrability as proposed by Caux and Mossel\cite{Caux:2011dx} is quite useful in this respect. Consider a given Hamiltonian on a lattice with $N$ sites in a preferred basis. Any operator $\mathcal{Q}$ expressed in this basis has a certain number of nonzero matrix elements. How the number of nonzero entries scale with system size (for example linear, polynomial, subexponential, or exponential) is called the 'density character' of that operator. A system is considered integrable if all IOM have at most a subexponential density character in this preferred basis. In other words, a system is integrable if the amount of information needed to specify the IOM is less than exponential in the system size.

Whereas Ref. \cite{Caux:2011dx} introduced their notion of integrability specifically for the real-space basis, one can naturally apply this same definition to the classical $\tau$-basis. In this basis, all the integrals of motion $\tauz_i$ have an extremely trivial matrix structure, and since they are essentially bits contain the smallest nontrivial amount of information. There is one IOM, however, that is not necessarily trivial, which is the Hamiltonian itself!

In the most general case the $\tau$-basis Hamiltonian contains of the order $2^N$ nonzero parameters $\xi_i, J_{ij}, J_{ijk}, $ etc. Following Ref. \cite{Caux:2011dx}, we postulate that a system is integrable if in the $\tau$-basis {\em at most subexponentially many Hamiltonian parameters are not vanishing}. All other matrix elements can be nonzero as long as they tend to zero in the thermodynamic limit $N \rightarrow \infty$.

Any noninteracting system trivially satisfies this condition, since $J_{ij}$ and higher order interaction parameters are all zero. It is interesting, however, to look at the ferromagnetic Heisenberg chain. The ground state is completely polarized, and starting from there the $\tau$-bits can represent the occupation of spin waves. Up to fourth order, the Hamiltonian in the $\tau$-basis reads
\begin{equation}
	H = \sum_k \epsilon_k \tauz_k + \sum_{k<k'} J_{kk'} \tauz_k \tauz_{k'}
	\label{H4th}
\end{equation}
where $\epsilon_k = J(1- \cos k)$ is the free spin wave dispersion. Using the Bethe ansatz (see for example \cite{1998cond.mat..9162K} for a pedagogical introduction), it is easy to show that the two-magnon spectrum contains $\sim N$ bound states and $\sim N^2$ so-called scattering states. The difference between the exact scattering states energies and two-magnon states vanishes when $N \rightarrow \infty$. The energy differences for the bound states, however, do not vanish, and consequently $\sim N$ of the $J_{ij}$ parameters will remain nonzero in the thermodynamic limit. Instead of $N^2$ nonzero Hamiltonian parameters $J_{ij}$ we only have $N$. This trend continues to three-magnon states, and higher, leading to a subexponential number of nonzero Hamiltonian parameters $J_{i_1 \ldots i_k}$. Since the Heisenberg chain is indeed integrable this result is consistent with our suggested definition of integrability.

Another way to characterize integrability is through the so-called {\em level statistics}. The Berry-Tabor conjecture\cite{Berry:1977bg,2015arXiv150906411D} states that the distribution of energy gaps $\delta_n = E_{n+1} - E_n$ between neighboring eigenstates in an integrable system is Poissonian, $P_g(\delta) \sim e^{-\delta}$. In non-integrable systems, in contrast, there is level repulsion so that the distribution goes to zero for zero energy gaps, $P_g(\delta \rightarrow 0) = 0$. 

For noninteracting theories it can be easily shown that $P_g(\delta \rightarrow 0) \neq 0$, consistent with the Berry-Tabor conjecture. The Hamiltonian can be written as $\hat{H}=\xi_i \hat{n}_i$, where the parameters $\xi_i$ are chosen from a random distribution $P( \left\{ \xi_1 \ldots \xi_L \right\} )$. The spectrum of {\em all} energy differences between many-body states equals
\begin{equation}
	F(\omega) = \int dt \, e^{i \omega t} \left| 2^{-L} \Tr \; e^{-i \hat{H}t} \right|^2
		= \int dt \, e^{i \omega t} \prod_{i=1}^L \cos^2 \left( \frac{\xi_i t}{2} \right)
\end{equation}
If there is level repulsion, this spectrum vanishes, $F(\omega) \rightarrow 0$, at small frequencies, $\omega \rightarrow 0$. Therefore $F(\omega=0) \neq 0$ would imply integrability, and we can compute the disorder-averaged spectrum
\begin{equation}
	\overline{F}(\omega) = \int d^L\xi \; P(\{ \xi_1 \ldots \xi_L \} ) \; F(\omega)
\end{equation}
to test this hypothesis. If the single-particle energies are independently distributed, $P(\{ \xi_1 \ldots \xi_L \} ) = \prod_i P_\xi (\xi_i)$, if follows that $\overline{F}(\omega)$ is the convolution of $L$ times the distribution $\widetilde{P}(\omega) = \int d\xi dt e^{i \omega t} P_\xi (\xi) \cos^2 \xi t$. The central limit theorem for large $L$ tells us that this distribution becomes normal, and since $\widetilde{P}(\omega)$ has mean zero it follows that $\overline{F}(\omega = 0 ) > 0$, thus proving that the system exhibits no level repulsion.

More general, this absence of level repulsion between many-body states comes from the fact that $2^L$ energy levels need to be constructed out of only $L$ single-particle energies. The above arguments can thus be generalized to the case where, for example, we look at the $\tau$-basis Hamiltonian $\hat{H}=\xi_i \tauz_i + J_{ij} \tauz_i \tauz_j$ cut-off at fourth order. The $\frac{1}{2} L (L+1)$ nonzero parameters do not provide enough freedom to allow for level repulsion between many-body states. Indeed, as long as there are less than exponential nonzero Hamiltonian parameters it is not possible to have level repulsion between the many-body energy levels. Our above postulate about integrability in the $\tau$-basis is thus consistent with the notion that integrable systems have no level repulsion.

\section{Numerical results on interacting disordered chains}
\label{Sec2Results}

In the previous section we have discussed general properties of the $\tau$-basis and presented an effective way of computing the parameters $J_{ij}$ in the effective Hamiltonian using few-particle exact diagonalization. We will now use this method to study the traditional model exhibiting many-body localization: the one-dimensional Anderson model with nearest neighbor interactions.

\subsection{The model}
We will consider the following $d=1$ dimensional model of spinless fermions,
\begin{equation}
	\hat{H} = -t \sum_{i=1}^{L-1} \left( \cd_{i} \c_{i+1} + h.c. \right)
		+ \sum_{i=1}^L \phi_i \hat{n}_i
		+ V \sum_{i=1}^{L-1} \hat{n}_i \hat{n}_{i+1}
\end{equation}
where $t=\frac{1}{2}$ is the nearest neighbor hopping, $\phi_i$ is a random onsite energy chosen from the uniform distribution $[-W,W]$, and $V=1$ is the nearest neighbor repulsion.   With these parameters, the model is equivalent to the Heisenberg chain with random field and nearest-neighbor coupling $J=1$. We choose open boundary conditions, and we consider various different lengths $L$ of the chain up to the largest value $L=60$. For most calculations we have considered values of the disorder from $W=1$ to $W=7$ in steps of $\Delta W = 1/4$.
For each data point we averaged over 1000 disorder realizations.

For a given disorder realization we compute the $\tau$-basis following  the method described in Sec. \ref{SecPractical} and Sec. \ref{SecNonUnique}.
Specifically, we use the few-particle diagonalization to extract the transformation $\hat{\mathcal{U}}_n$ in the corresponding state space and extend this result to many-body states following Eqns. (\ref{U4Eqn})-(\ref{U4Eqn2}).
We end up with an operator $\hat{U}_{2n}$ that diagonalizes the Hamiltonian up to a given order and then rotate any other operator of interest with the same $\hat{U}_{2n}$. 
In particular, we transform the density operator $\hat{n}_j$ to obtain $\tauz_j$ and the operator $\cd_{i} \c_{j}$ to quantify  the localization properties of many-body states.

\subsection{Comparison with exact diagonalization}

Let us start with a benchmark of the accuracy of the method. We therefore consider small systems that can be solved using exact diagonalization of the full spectrum. The exact ground state energy for half-filled states is then compared to the ground state energy of the classical Hamiltonian up to 4th order, Eq. (\ref{H4th}), or up to 6th order for all $n$-particle states of the form 
\begin{equation}
|\Phi_n\rangle = \taud_{i_1}\taud_{i_2}\cdots \taud_{i_n}|0\rangle
\label{state}
\end{equation}
where $\taud_i$ is the creation operator of a particle at state $i$ in the classical basis up to 4th or 6th order.
In this basis, the expectation value of the energy is just the sum of the single particle energies of the  $n$ occupied states in Eqn.~(\ref{state}), the $J_{ij}$ terms for the $n(n-1)/2$ occupied pairs of states, and if we include the 6th order terms also the $J_{ijk}$ terms.

\begin{figure}
	\includegraphics[width=\columnwidth]{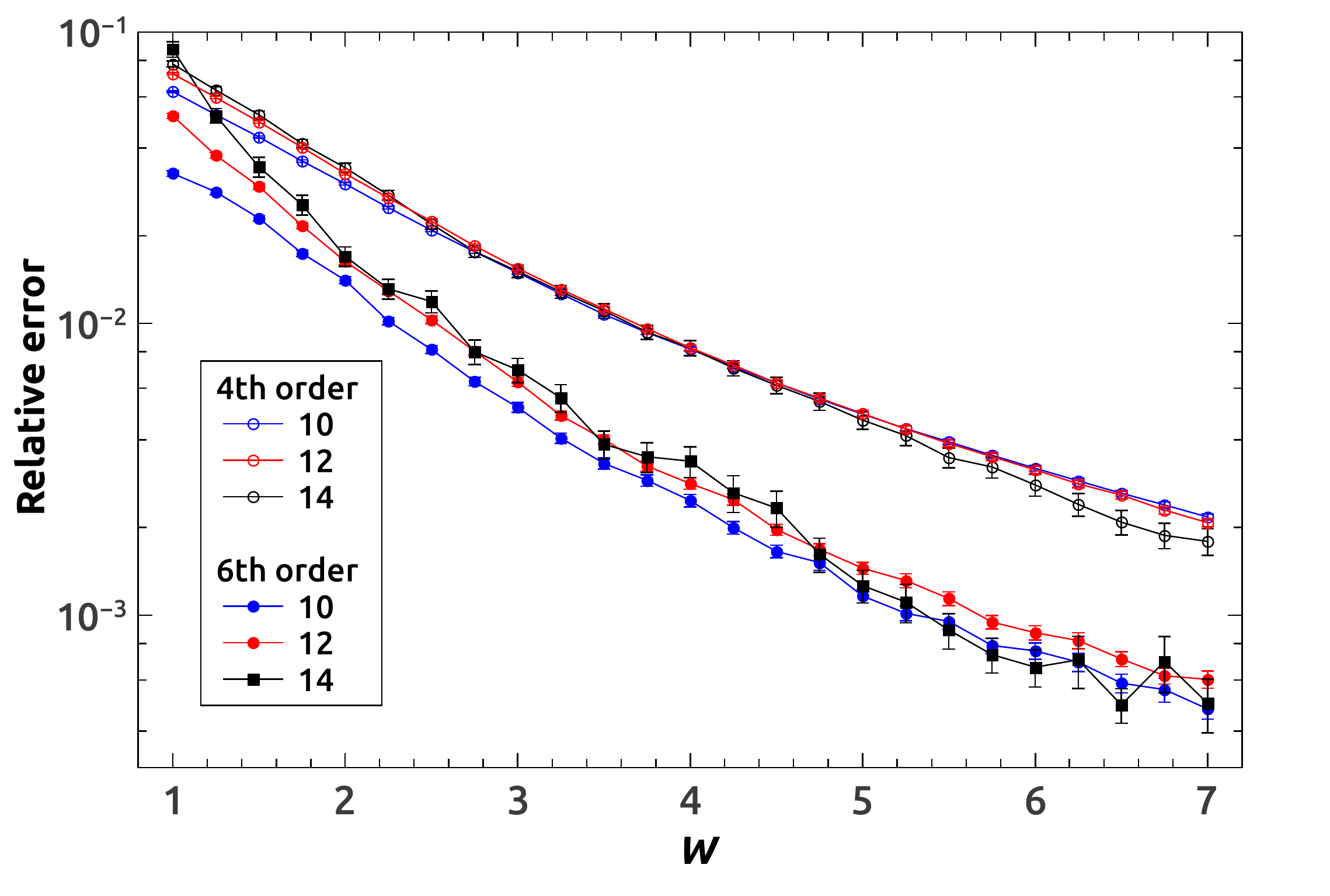}
	\caption{Relative error in the ground state energy as a function of disorder for several system lengths with half-filling occupation. The energy is estimated from the coefficients of the Hamiltonian up to fourth (empty symbols) and sith (solid symbols) orders.\label{FigEDcomparison}}
\end{figure}

In Figure \ref{FigEDcomparison} we plot the relative error of the ground state energy at half-filling. The upper set of curves (empty symbols) corresponds to 4th order, while the lower set (full symbols) to 6th order. 
The method is more adequate at large disorder $W$, but works fairly well in the whole range studied. It is able to predict the ground state energy of a large finite density state from information obtained by diagonalizing few particles systems.

What is important to realize that for all disorder values $W>1$ the accuracy of the method is increased by going to higher order in fermion operators. This suggests that our systematic order-by-order expansion introduced in the context of displacement transformations\cite{Rademaker:2016jf} is valid. For very small disorders $W \leq 1$ the expansion seems to break down, but this might change if one includes higher order terms. For now we will focus on the disorder values where the expansion is convergent.

\subsection{Spread of the integrals of motion}

\begin{figure}
  \begin{center}		
	\includegraphics[width=\columnwidth]{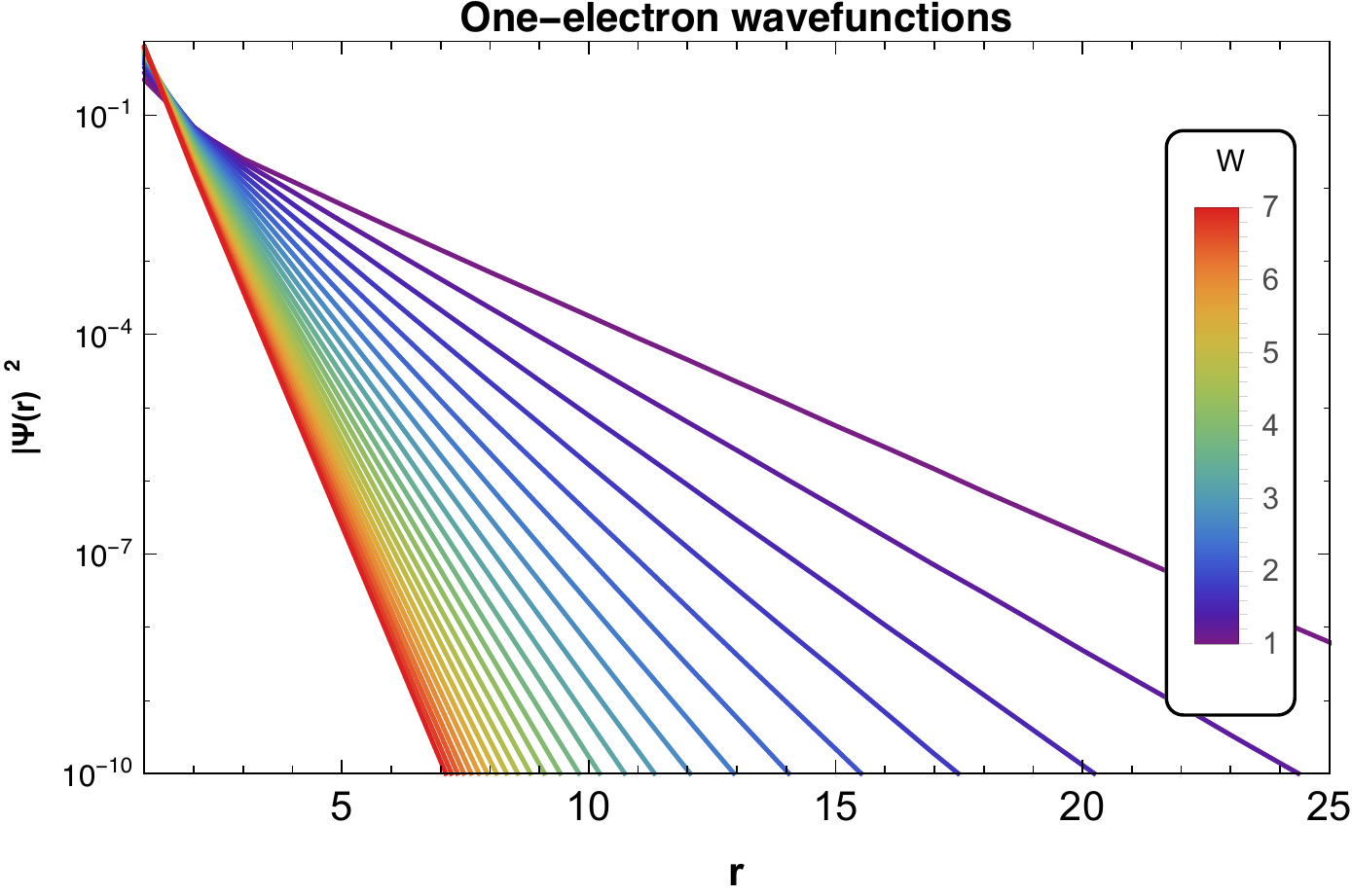}
	\includegraphics[width=\columnwidth]{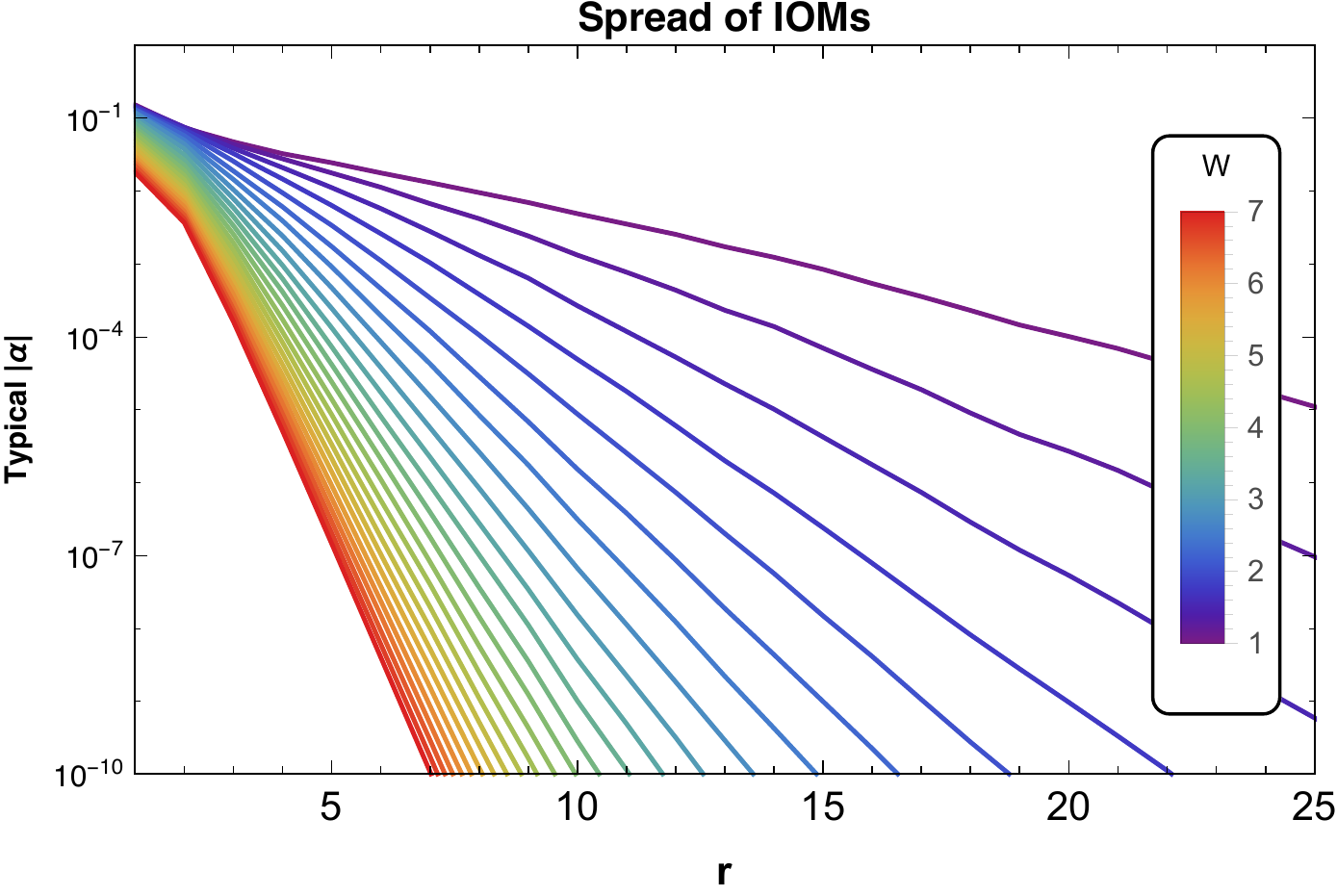}
  \end{center}
	\caption{Spread of the IOM for various disorder strengths. Top: the square of the one-electron wavefunction as a function of distance. Bottom: Upon inclusion of interactions, we can measure the weight of each IOM following Eqn. (\ref{TauBasisExpansion}).
	\label{FigIOM}}
\end{figure}

Each IOM $\tauz_{i_0}$ can be expanded in terms of operators in the original basis, as in Eqn. (\ref{TauBasisExpansion}).
We first consider how the diagonal terms (those only involving density operators $\hat{n}_i$) in this expansion decay with distance.
In the bottom panel of Figure \ref{FigIOM} we plot the median of the absolute value of these terms $|a_{i;jk}|$ and $|\alpha_{i; jklm}|$ up to 4th order on a logarithmic scale as a function of $\max(|i_0-i|,|i_0-j|)$ for several values of the disorder and for a system size $L=60$. We note an exponential decay with distance of the components of the IOM for all values of the disordered considered.
For comparison, we also plot in the top panel the distance dependence of the modulus square of one-electron wavefunctions. It is clear that the inclusion of interactions increases the localization length. Later we will quantitatively compare all the different localization lengths.

\begin{figure}
    \includegraphics[width=\columnwidth]{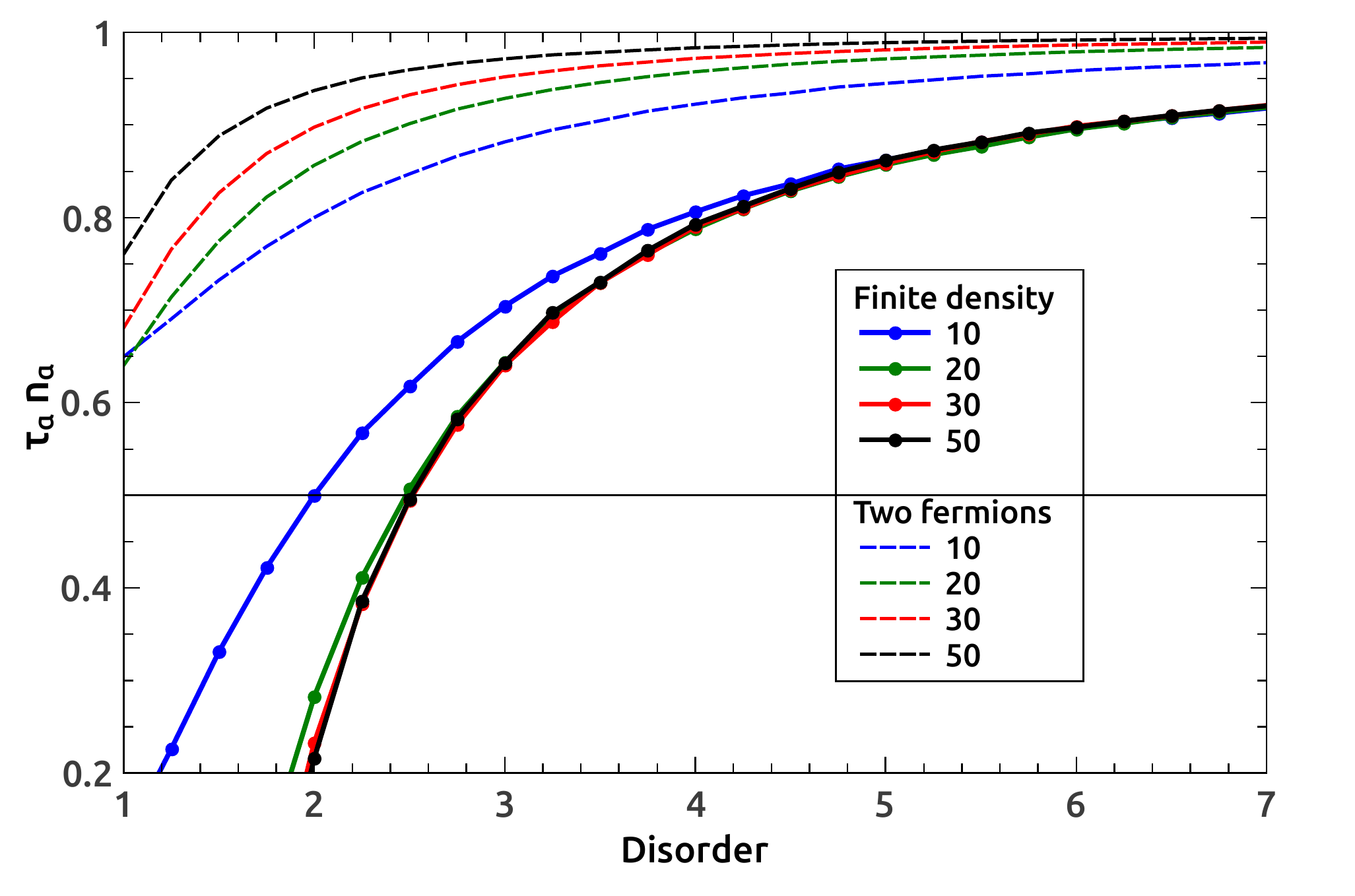}
	\caption{Average overlap between an IOM and its original density operator as a function of disorder for several system sizes. Solid lines correspond to traces over finite density states, while dashed lines to traces over two particle states.\label{Foverlap}}
\end{figure}

A different way to quantify the spread of the IOM is through the overlap
\begin{equation}
	O(i,j) = \frac{ \Tr \; \tauz_i \hat{n}_j }{\Tr \; \hat{n}_j }.
\label{overlap}
\end{equation}
At the same site, it should return $O(i,i) =1$ if the integral of motion $\tauz_i$ is completely localized at site $i$. If $\tauz_i$ is completely delocalized, the overlap should reduce to $\frac{1}{2}$ (which can be easily seen by computing $\Tr \; \hat{n}_i \hat{n}_j  / \Tr \hat{n}_i= \frac{1}{2}$ for $i \neq j$).

We also need to specify over which states we perform the trace in Eqn. (\ref{overlap}). Obviously, the trace depends on the number of particles of the state space.
The trace of any operator $\hat{O}$ in the subspace of a fixed number of particles $k$ can be constructed from fewer-particle traces. Let us assume that the diagonal part of the operator $\hat{O}$ is split in terms of the number of density operators involved as,
\begin{equation}
 \hat{O}=\hat{O}_1+\hat{O}_2+\hat{O}_3+\cdots
\end{equation}
where $\hat{O}_1$ only contains terms with only one density operator, $\hat{O}_2$ only contains terms that are the product of two density operators, etc. The trace over the state space of $k$ particles in $L$ sites is
\begin{equation}
\frac{\Tr \hat{O}}{\cal N}
	=\frac{k}{L}\Tr_1\hat{O}_1
	+\frac{k(k-1)}{L(L-1)}\Tr_2\hat{O}_2
	+\frac{k(k-1)(k-2)}{L(L-1)(L-2)}\Tr_3\hat{O}_3+\cdots
\end{equation}
where $\Tr_n$ is the trace over the subspace of $n$ particles, and ${\cal N}$ is the total number of states.

To measure the degree of delocalization produced by the interactions, we consider in Eq.\ (\ref{overlap}) the density operator for the one-particle state $\alpha_i$ that has the most overlap with the original site state $i$.
We have computed $O(i,\alpha_i)$ as a function of disorder, which is shown in Figure \ref{Foverlap}.
The solid curves correspond to the trace performed in half-filled $k=L/2$ systems, while dashed lines correspond to two-particle systems. It is clear that our method captures the drastic difference between finite density states and few particle states, even though we only diagonalized few-particle states. For two electron states, the overlap tends to 1 for all disorder as system size increases, since eventually the two electrons will not see each other. 
At half-filling, the curves quickly tend to a size-independent behavior which reaches the value 1/2 for disorders slightly smaller than 3. 

Notice that for disorder $W<3$ the trace seems to have unphysical values less than $1/2$. This is due to our cut-off of the unitary transformation $\hat{U}$ at 4th order. Inclusion of terms at higher order will make the total curve lie in the region between $1/2$ and $1$.

\begin{figure}
  \begin{center}		
	\includegraphics[width=\columnwidth]{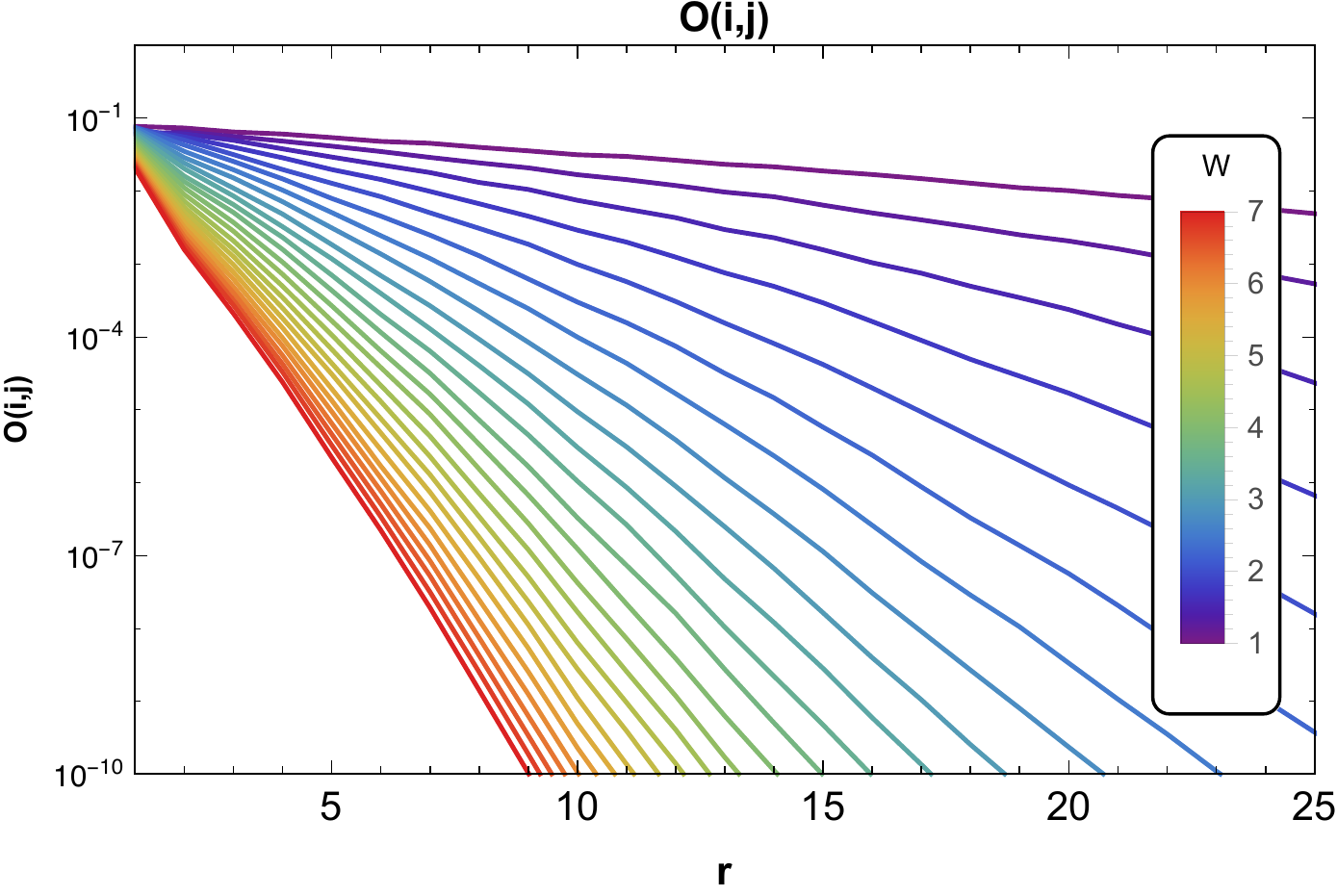}
  \end{center}
	\caption{Spatial dependence of $O(i,j)$ according to Eqn. (\ref{overlap}) for various disorder strengths. 
	\label{FigIOM2}}
\end{figure}

\begin{figure}
  \begin{center}		
	\includegraphics[width=\columnwidth]{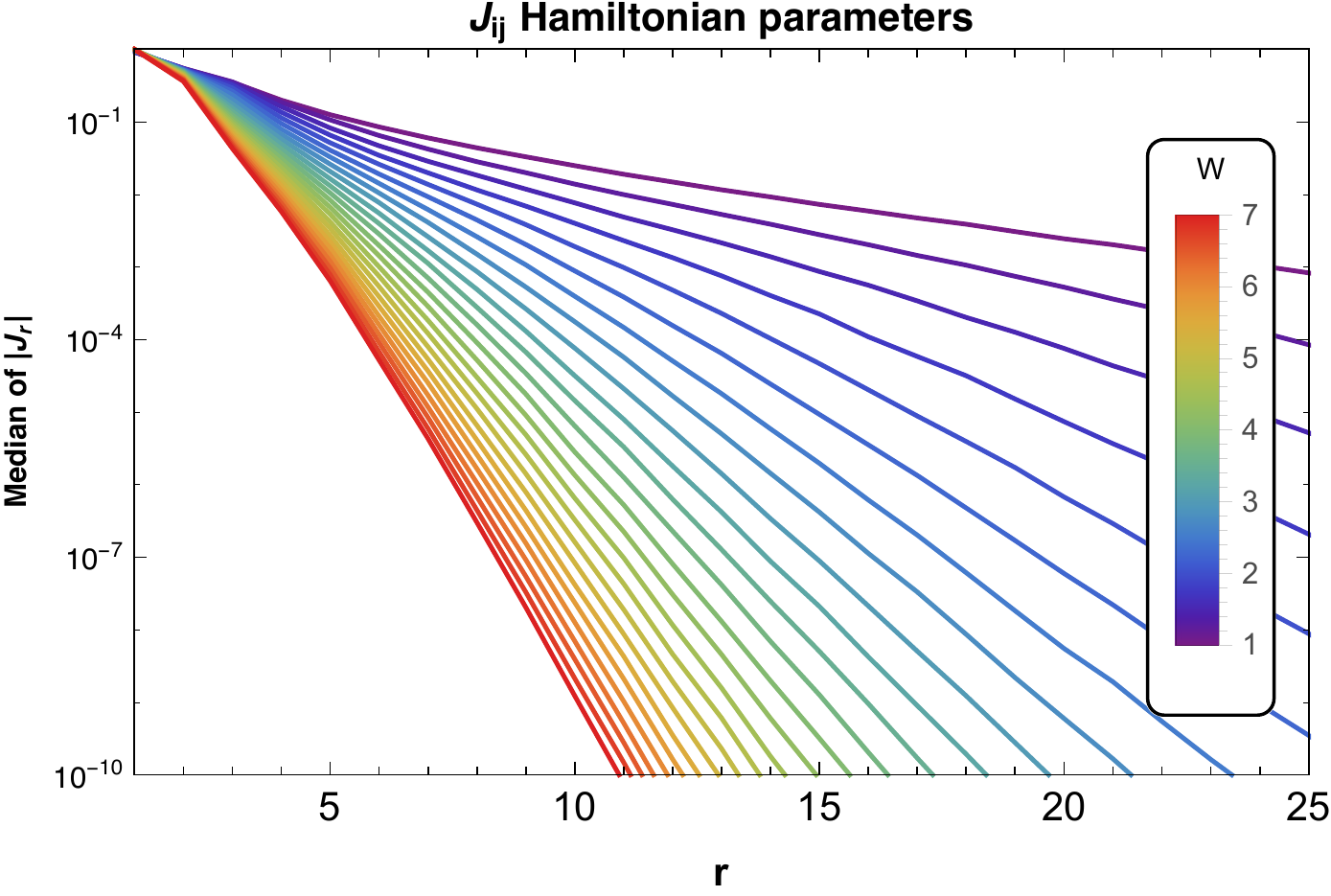}
  \end{center}
	\caption{Spatial dependence of the typical value of $|J_{ij}|$ as a function of distance $|i-j|$ for various disorder strengths $W$.
	\label{FigIOM3}}
\end{figure}

\begin{figure}
  \begin{center}		
	\includegraphics[width=\columnwidth]{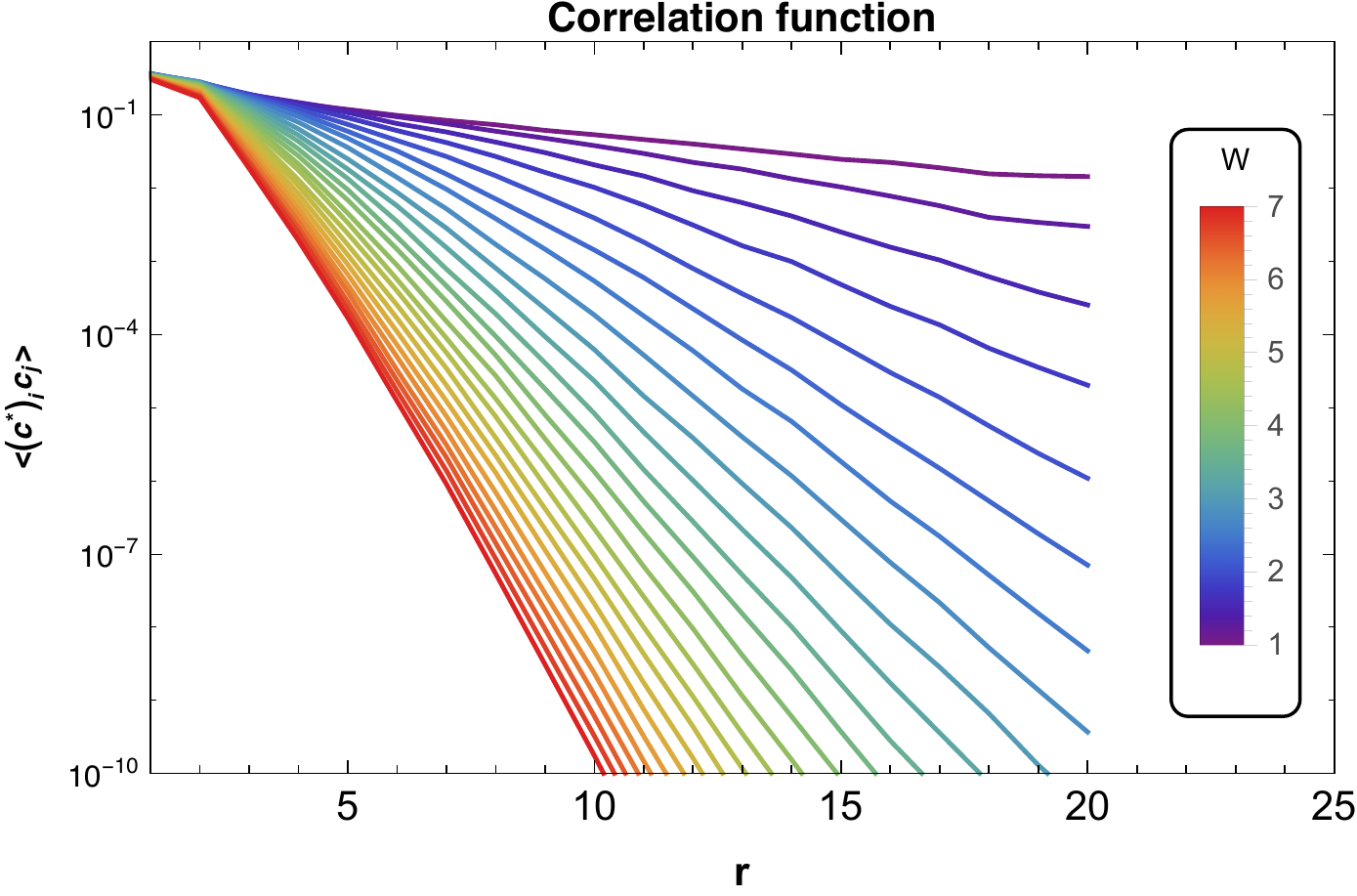}
  \end{center}
	\caption{Spatial dependence of the correlation function Eqn.~(\ref{corre}) for various disorder strengths $W$.
	\label{FigIOM4}}
\end{figure}

We have also computed the decay of $O(i,j)$ as a function of $|i-j|$. We look at  the geometric mean over many disorder realizations. The results are plotted in Figure \ref{FigIOM2} on a logarithmic scale for various disorder strengths.

\subsection{Effective interactions between IOMs}
\label{SecIOMmedian}

Subsequently we look at the decay of the Hamiltonian parameters $J_{ij}$ with distance. In Figure~\ref{FigIOM3} we plot the median of  $|J_{ij}|$ as a function of distance ($|i-j|$) for the same values of the disorder and system size $L=60$ as mentioned above. For all values of the disorder the terms $|J_{ij}|$ show a particularly good exponential decay with distance.

This is also a point to observe that the sum of all diagonal parameters $J_{i_1 \ldots i_k}$ remains constant under our flow. That is, for our nearest neighbor interacting system we find
\begin{eqnarray}
	\sum_{i=1}^L \xi_i &=&  \sum_{i=1}^L \phi_i, \\
	\sum_{1 \leq i<j\leq L}  J_{ij} &=& V(L-1), \\
	\sum_{1 \leq i<j<k\leq L}  J_{ijk} &=& 0,
\end{eqnarray}
and so forth for higher order terms.

Interestingly, up to 4th order the typical $J_{ij}$ parameters do not display any qualitative change when going from the MBL phase $W \gtrsim 3.5$ to the ergodic phase at $W \lesssim 3.5$. However, possible rare fluctuations of $J_{ij}$ are neglected by looking at the typical value of $J_{ij}$. In Sec. \ref{SecDistr} we therefore study the full distribution of $J_{ij}$.

\subsection{Correlation function}

Our final measure of localization is by studying the distance dependence of the correlation function $\langle \cd_{i} \c_{j}+\cd_{j} \c_{i}\rangle$, defined as
\begin{equation}
\langle \cd_{i} \c_{j}\rangle=\left\langle \ln \sum_{\alpha} \left|\langle \Psi_\alpha |( \cd_{i} \c_{j}+\cd_{j} \c_{i})|\Psi_\alpha\rangle\right|^2 \right\rangle_{\rm disorder}
\label{corre}.
\end{equation}
This quantity is not a trace and so it is much more difficult to calculate than for example Eq.\ (\ref{overlap}). It is not possible to obtain exactly its expectation value for finite density states in terms of the expectation values for few particle states.
As the contributions to the sum are very widely distributed, we approximated this by the maximum contribution, which allows us to determine straightforwardly the final density result. 
The results are shown in Figure \ref{FigIOM4}. In this case the system size is $L=40$. Again the data displays an exponential dependence with distance and the overall behavior is similar to the rest of the quantities considered, albeit with a different localization length.

\begin{figure}
  \includegraphics[width=\columnwidth]{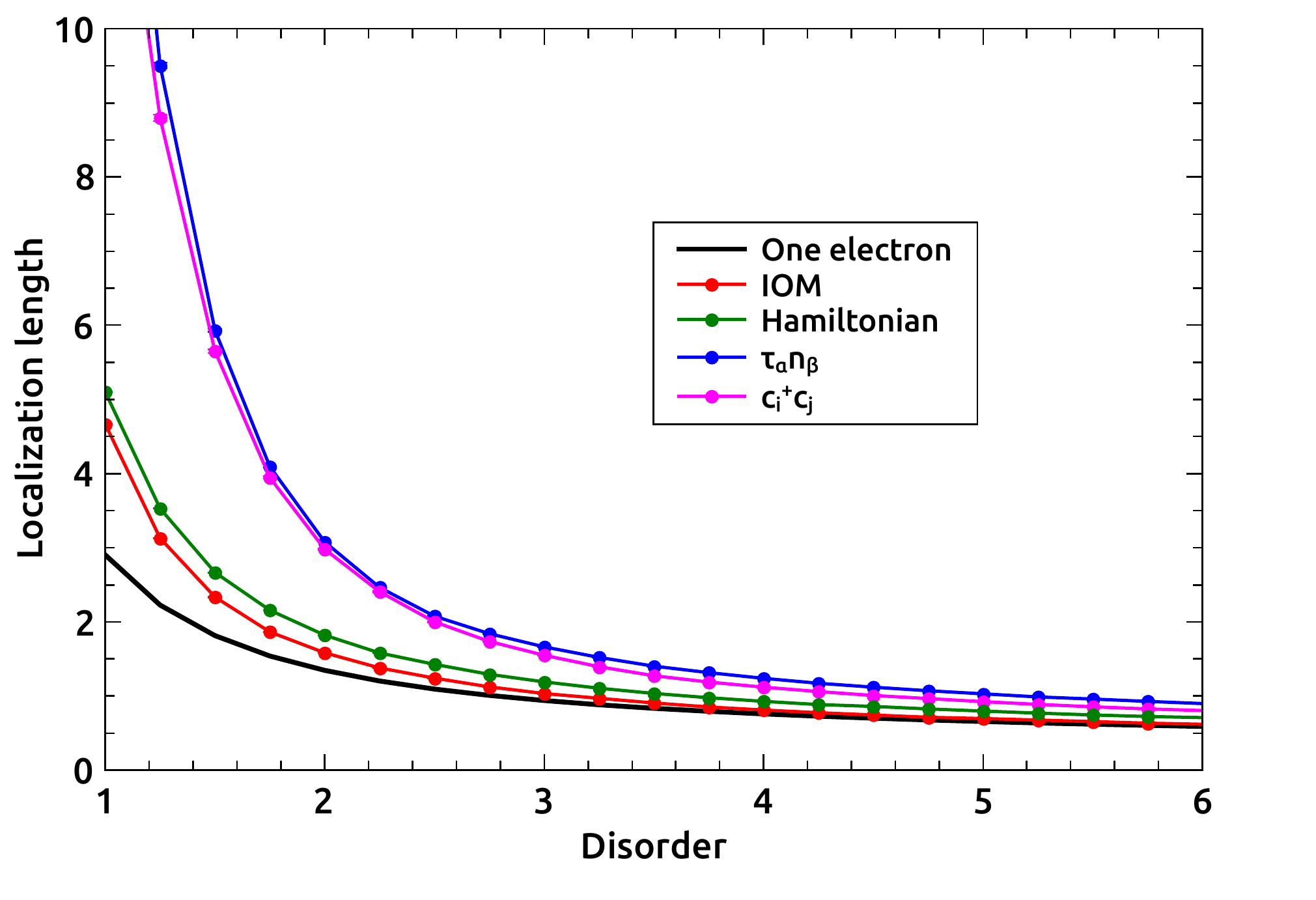}
	\caption{Localization lengths as a function of disorder for the magnitudes indicated in the figure. See Sec. \ref{SecLocalL} for a comparison of the different lengths.}
	\label{FigLoca}
\end{figure}

\begin{figure*}
 \begin{center}
  \includegraphics[width=\columnwidth]{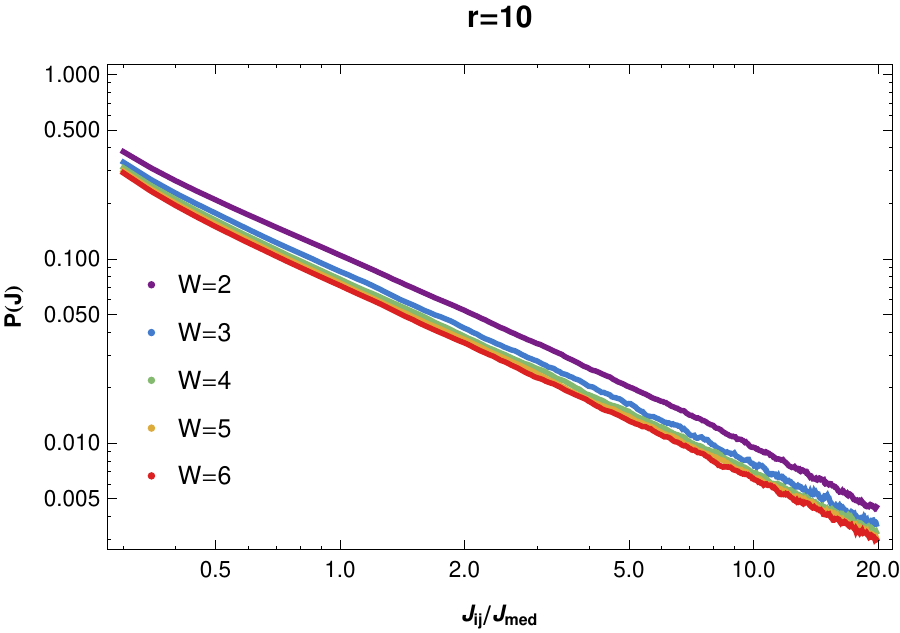}
  \includegraphics[width=\columnwidth]{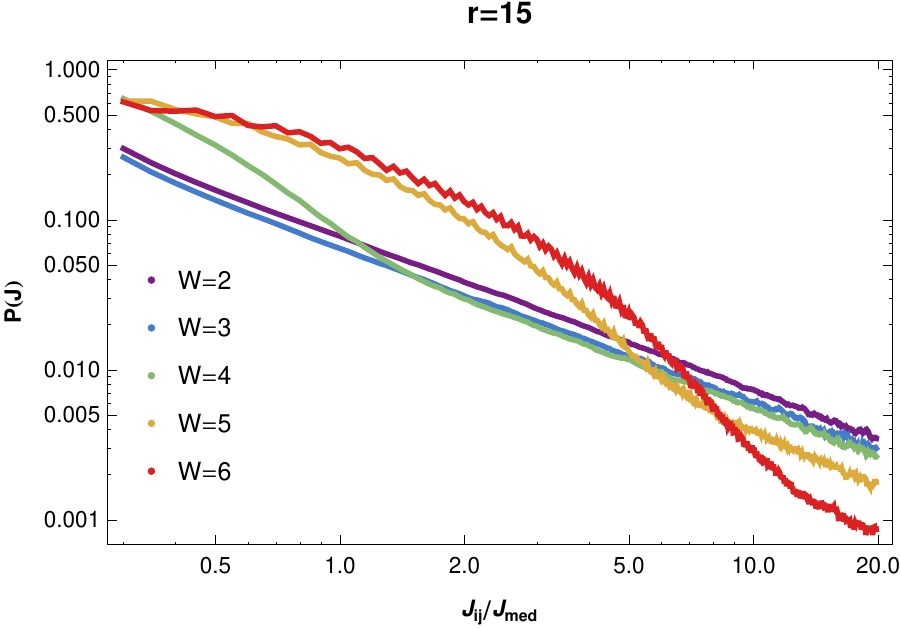}
  \includegraphics[width=\columnwidth]{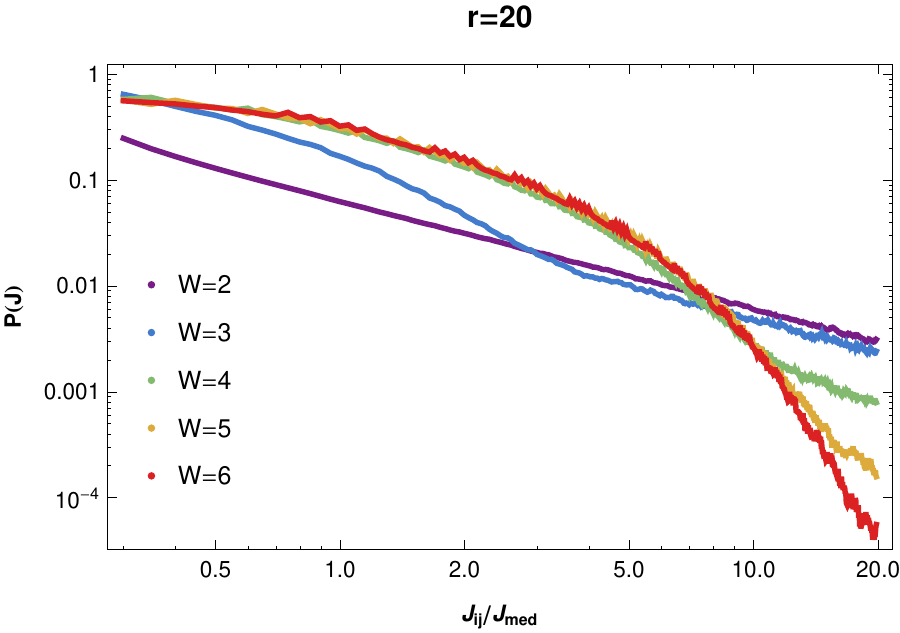}
  \includegraphics[width=\columnwidth]{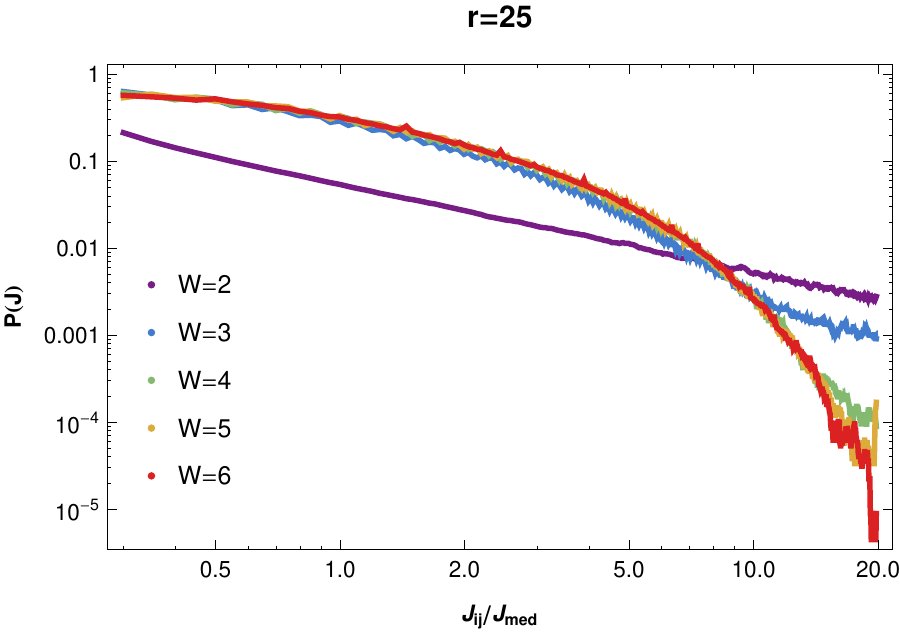}
 \end{center}
  \caption{The distribution $P(|J|)$ of $J_{ij}$ parameters for various distances and disorder strengths, in a log-log plot. We normalized the horizontal axis by the median value $J_{\mathrm{med}}$.}
  \label{FigDistribution}
\end{figure*}

\subsection{Localization lengths}
\label{SecLocalL}

All quantities studied so far have shown a distance dependence roughly exponential and so it is natural to define a localization length $\xi$ for each of them through the expression
\begin{equation}
-\frac{2|i-j|}{\xi}\propto \ln A(i,j)
\label{loca}
\end{equation}
where $A(i,j)$ refers generically to any of the quantities previously studied.
The factor of 2 in Eq.\ (\ref{loca}) is to accommodate to the standard definition of the localization length for one-electron systems. 

In Figure \ref{FigLoca} we represent the localization lengths for the coefficients of the IOM (red), for the overlap $O(i,j)$ (blue), for the coefficients of the Hamiltonian $J_{ij}$ (green) and for the correlation function 
$\langle \cd_{i} \c_{j}\rangle$ (magenta). All these quantities have been calculated up to 4th order. The black curve corresponds to the one-electron localization length.
One can appreciate that the coefficients of the IOM and of the Hamiltonian present very similar localization lengths, quite close to the one-electron values at large disorders and getting increasingly larger as the disorder decreases.

The localization lengths for $O(i,j)$ and for $\langle \cd_{i} \c_{j}\rangle$ are very similar to each other and are much larger than the rest. Recall that these two quantities are obtained from expectation values for finite density states. If instead, they would have obtained from expectation values for two electron states, localization lengths much more similar to one-electron values would have been produced.

It is remarkable that the properties of the IOM do not display any signature of the delocalization transition at this level, even though the computed ground state energies are relatively accurate (see Figure \ref{FigEDcomparison}). There are two possible reasons for this behavior, which we will briefly touch upon in the last two subsections.

\subsection{Distribution of $J_{ij}$ parameters}
\label{SecDistr}

As mentioned in Sec. \ref{SecIOMmedian} the typical value of $J_{ij}$ displayed exponential decay as a function of $|i-j|$. However, in disordered systems it is natural to expect that rare fluctuations play a relevant role. To answer this question, we studied the full distribution of the $|J_{ij}|$ Hamiltonian parameters, both as a function of distance and as a function of disorder. Our results are shown in Figure \ref{FigDistribution} for four distances $r=10,15,20$ and $r=25$ for 5 different values of disorder. The horizontal axis is normalized by the median value of $J_{ij}$.

\begin{figure}
	\begin{center}		
		\includegraphics[width=\columnwidth]{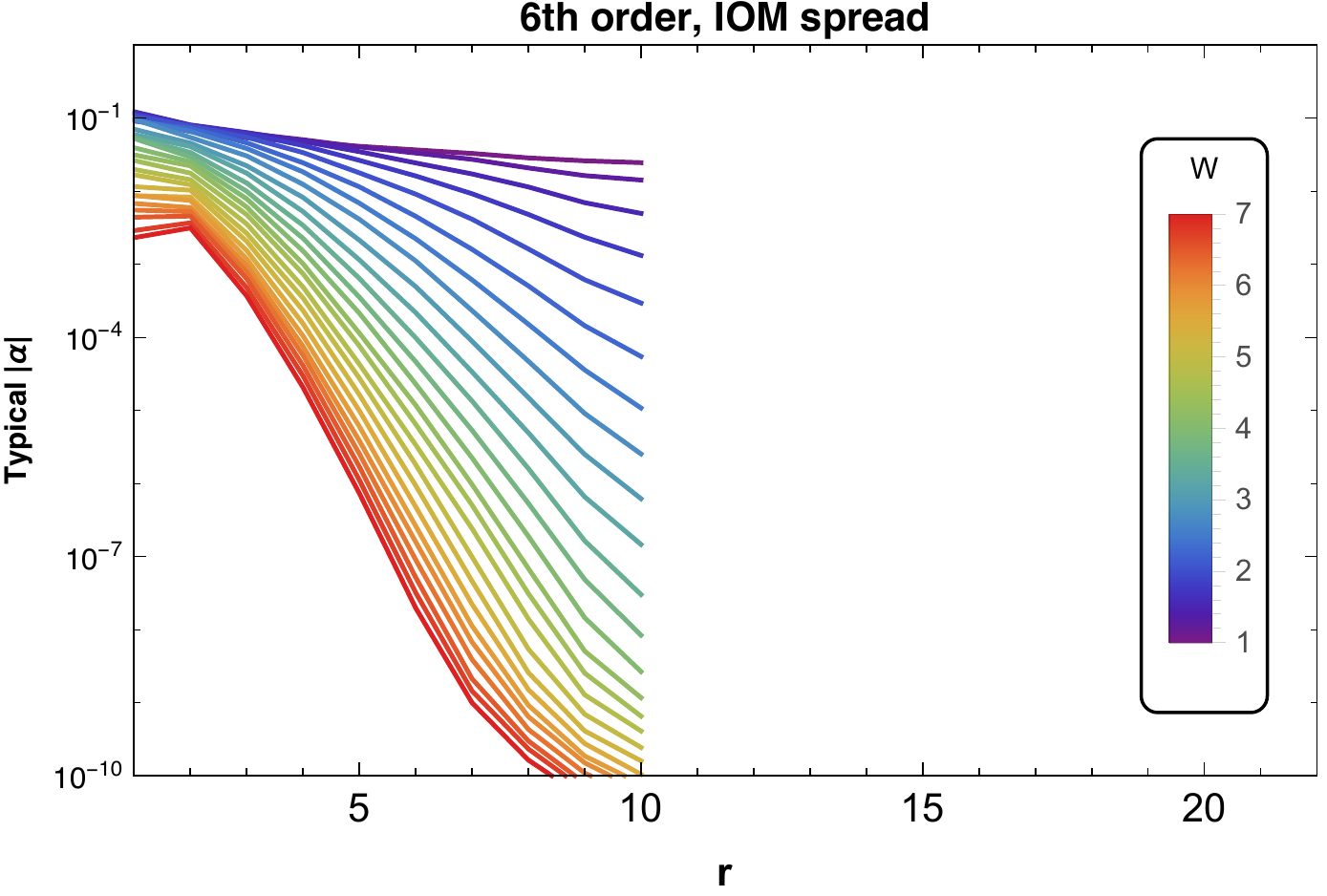}
		\includegraphics[width=\columnwidth]{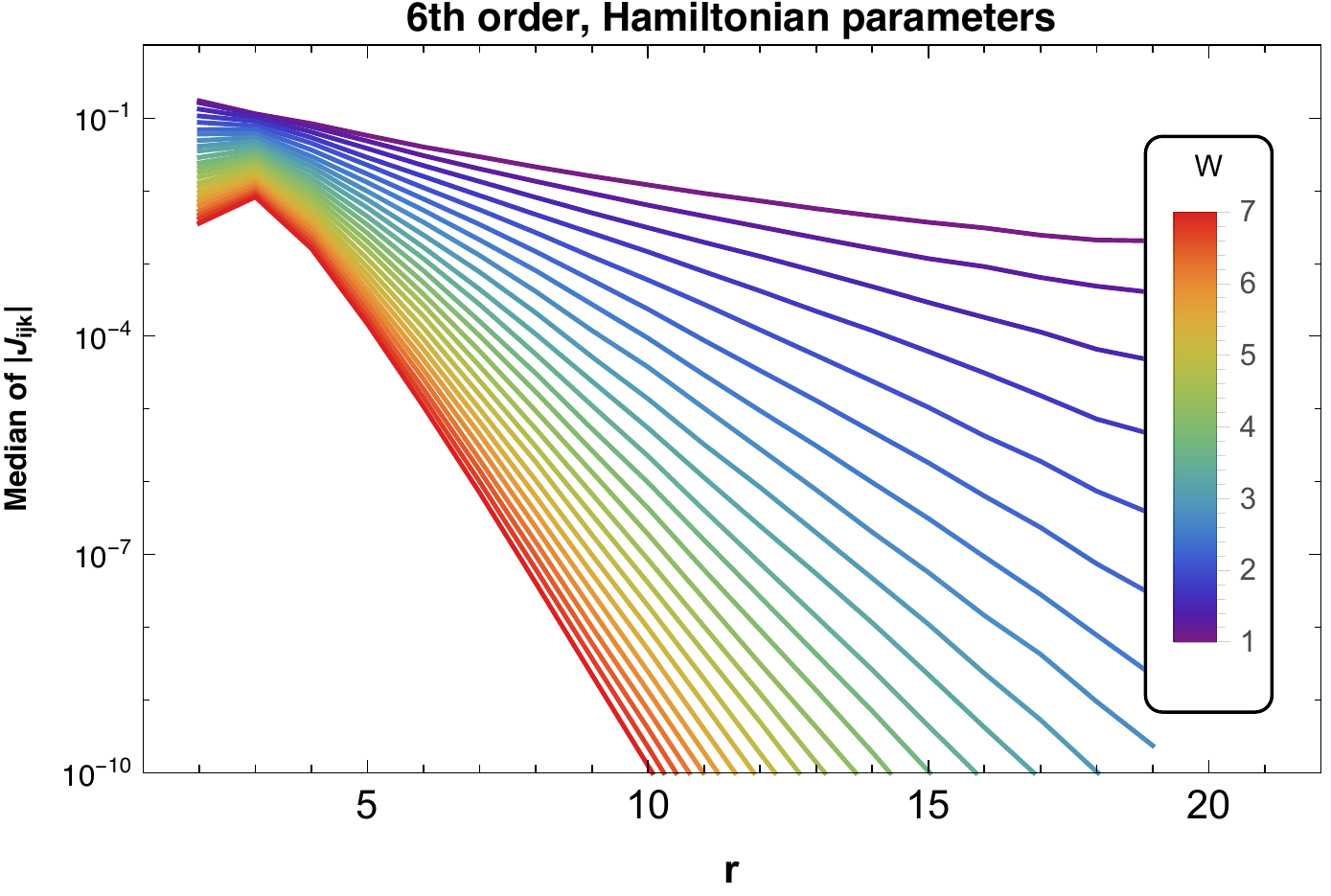}
	\end{center}
	\caption{Spatial dependence of the median of the coefficients of the IOMs (top panel)  and of the Hamiltonian parameters $J_{ijk}$ (bottom panel) as a function of distance for various disorder strengths.
		\label{Fig6th}}
\end{figure}

In Ref. \cite{2016arXiv160707884P} it was claimed that the MBL phase is characterized by a $1/f$ distribution, while in the metallic phase the distribution tends to a constant at the small values tail. We see different behavior, however. At strong disorder we see that that the distribution is exponentially decaying for large $|J|$ at large distances. Going to shorter distances and/or weaker disorder, there is a shifting crossover point to a $1/J$ distribution. In particular, for the weakest disorder $W=2$ the distribution seems to be independent of distance and always a $1/J$ distribution. A $1/J$ tail implies that there are relatively many rare fluctuations and that the average value of $|J_{ij}|$ is ill-defined.

It is clear that for weak disorder rare disorder fluctuations become most relevant, given the stability of the $1/J$ distribution independent of distance. However, it is an open question whether these rare fluctuations not captured by typical values are sufficient to drive the system into an ergodic phase.

\subsection{Up to 6th order}

Another possible route to delocalization lies in the structure of IOM at higher order. We end our numerical results section by presenting partial results for calculations up to 6th order. In Figure \ref{Fig6th} we plot the typical coefficients of the IOM (top panel) and of the Hamiltonian $|J_{ijk}|$ (bottom panel) on a logarithmic scale as a function of distance for several values of the disorder.
As distance we take the maximum separation between the indices, which is $|i-k|$.
The system size is $L=20$. We still find that the decay is basically exponential, even for the lowest values of the disorder.

\begin{figure}
	\includegraphics[width=\columnwidth]{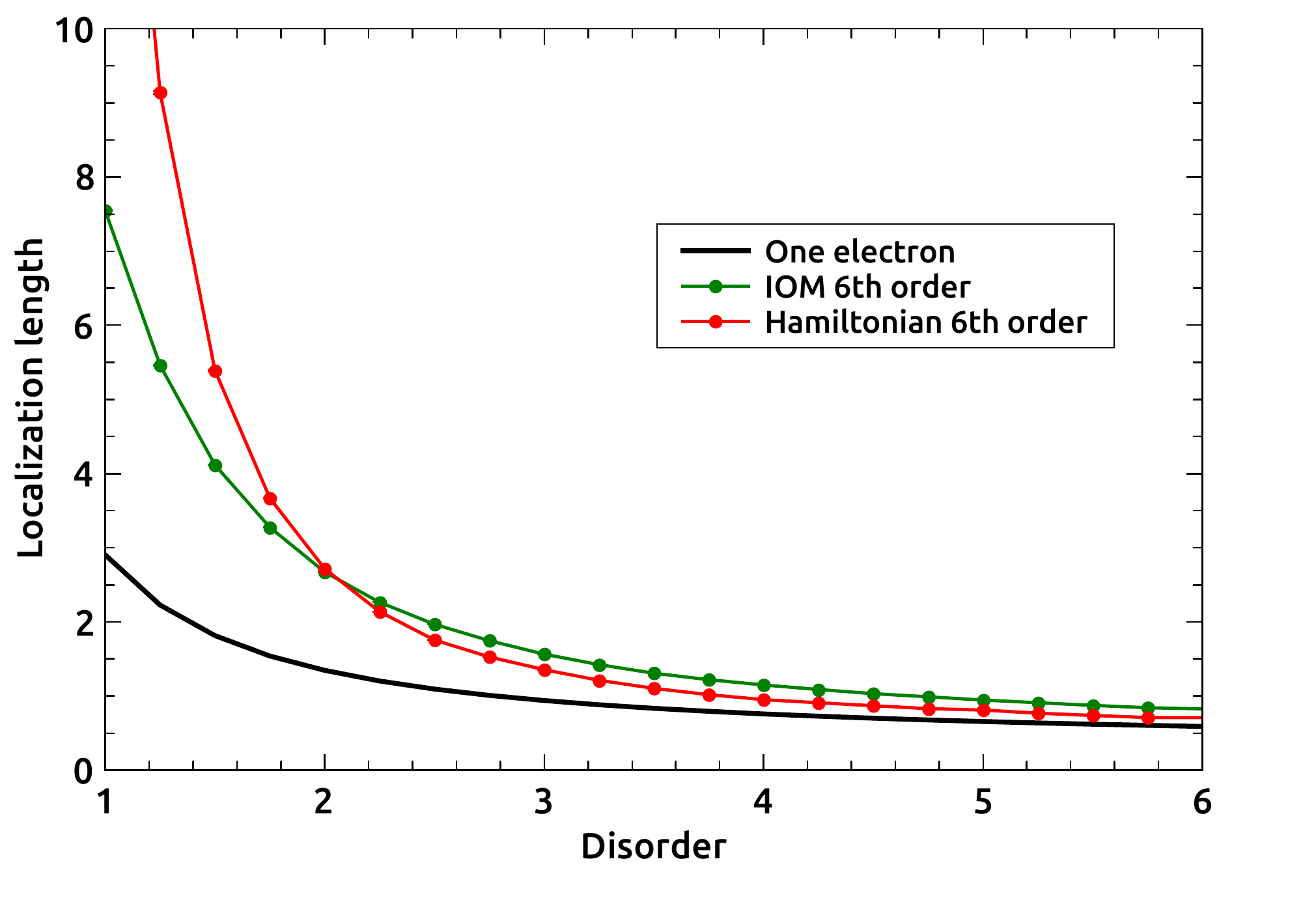}
	\caption{Localization length as a function of disorder for the IOMs and  the  Hamiltonian, when we include terms up to 6th order. For comparison, we show the one-electron localization length.}
	\label{FigLoca3}
\end{figure}

From the slopes of the curves in Figure \ref{Fig6th} we obtain effective localization lengths for the 6th order terms of the IOM and the Hamiltonian.
The results are shown in Figure \ref{FigLoca3} as a function of disorder.
The one-electron result has been included for comparison. Both localization lengths for the 6th order terms are larger than the similar ones for 4th order term, but they are smaller than the system size for all values of the disorder analyzed.

\section{Conclusions and outlook}

We have demonstrated that we can efficiently compute IOM and the corresponding effective Hamiltonian following Eqn. (\ref{LIOM}) up to 4th and 6th order in interacting disordered systems.

Deep in the many-body localized phase, for strong disorder $W$, the results are consistent with expectations of a fully MBL phase.\cite{2014arXiv1407.8480C,Huse:2014co,Serbyn:2013cl} The IOM are localized, and the interactions between them decay exponentially with distance. Surprisingly, the typical IOM remain localized throughout the phase diagram, even in regimes where exact diagonalization suggests\cite{Pal:2010gr} an ergodic phase. We present five different measures of the localization lengths (Figure \ref{FigLoca}). The localization lengths obtained from either the overlap of IOM with single-particle density operators $O(i,j)$, or the correlation function $\langle \cd_i \c_j \rangle$, are the same. We propose that these are the relevant typical many-body localization length characterizing the disordered system.

Our current results suggest that the role of rare fluctuations might be the key towards understanding the delocalization transition, similar to the infinite randomness fixed point.\cite{1995PhRvB..51.6411F,1994PhRvB..50.3799F,1992PhRvL..69..534F} For weak disorder, the interactions between IOM are distributed as $P(|J|) \sim 1/J$ independent of distance, showing that at every distance strong resonances exist. It would be interesting to see whether these rare fluctuations in these typically localized IOM can lead to ergodic behavior as observed in dynamical processes, such as the entanglement growth after a quench\cite{Bardarson:2012gc,Luitz:2016ez,2015arXiv151109144L} or the evolution of an initial density imbalance.\cite{Luschen:2016}

Under the assumption that the delocalization transition is a second order transition subject to a diverging correlation length, several critical properties have been suggested such as volume law entanglement at the transition\cite{2014arXiv1405.1471G} and the existence of a 'quantum critical fan' for finite size systems.\cite{PotterVasseurSid2015} However, as can be seen in Figure \ref{Foverlap}, we do not see significant size dependence of the localization of the IOM up to 4th order we considered. 

So even though we have proven that in general the $\tau$-basis can be constructed, it is nontrivial to extract as of yet the relevant phase diagram. In the case of the delocalization transition this can be resolved by either going to higher order or to more systematically study the rare fluctuations. Another likely fruitful approach is to combine displacement transformations with Hartree-Fock methods, which will provide the approximate IOM with respect to some finite density state, much akin to the traditional Fermi liquid theory energy functional.

\begin{acknowledgements}
L.R. was supported by the Dutch Science Foundation (NWO) through a Rubicon grant and by the National Science Foundation under Grant No. NSF PHY-1125915. 
M.O. and A.M.S. were supported by Spanish MINECO and FEDER (EU) Grant No. FIS2015-67844-R  and by the Fundacion Seneca FEDER (EU)  Grant 19907/GERM/15.
\end{acknowledgements}


\providecommand{\WileyBibTextsc}{}
\let\textsc\WileyBibTextsc
\providecommand{\othercit}{}
\providecommand{\jr}[1]{#1}
\providecommand{\etal}{~et~al.}

\end{document}